\documentclass[lettersize,journal]{IEEEtran}

\usepackage[english]{babel}
\usepackage{tikz}
\usepackage{amsmath}
\usepackage{algorithm}
\usepackage{algpseudocode}
\usepackage{tabularx}
\usepackage{multirow}
\usepackage[caption=false,font=footnotesize]{subfig}
\usepackage{url}
\usepackage{enumitem}
\usepackage{xcolor}
\usepackage[colorinlistoftodos, prependcaption]{todonotes}
\usepackage[most]{tcolorbox}
\newtcolorbox{highlight}{breakable, enhanced,
  colback=yellow!18, colframe=yellow!18, 
  boxrule=0pt,            
  boxsep=0pt,             
  left=0pt, right=0pt,    
  top=0pt, bottom=0pt,    
  sharp corners, arc=0pt, 
  width=\linewidth,       
  before skip=0pt, after skip=0pt 
}

\newcommand{\tom}[1]{\textcolor{black}{#1}}

\begin{document}

\title{SMaRTT: Sender-based Marked Rapidly-adapting Trimmed \& Timed Transport}

\author{%
Tommaso Bonato,~\IEEEmembership{ETH Zurich \& Microsoft},~
Abdul Kabbani,~\IEEEmembership{Microsoft},~
Ahmad Ghalayini,~\IEEEmembership{Microsoft},~
Anup Agarwal,~\IEEEmembership{Carnegie Mellon University},~
Daniele De Sensi,~\IEEEmembership{Sapienza University of Rome},~
Rong Pan,~\IEEEmembership{AMD},~
Costin Raiciu,~\IEEEmembership{University Politehnica of Bucharest \& Broadcom Inc.},~
Mark Handley,~\IEEEmembership{University College London},~
Mihai Brodschi,~\IEEEmembership{Broadcom Inc.},~
Timo Schneider,~\IEEEmembership{ETH Zurich},~
Nils Blach,~\IEEEmembership{ETH Zurich},
Daniel Santos Ferreira Alves,~\IEEEmembership{Microsoft},~
Torsten Hoefler,~\IEEEmembership{ETH Zurich \& Microsoft}%
}



\maketitle

\begin{abstract}
With the rapid growth of artificial intelligence (AI) workloads in datacenters, the Ultra Ethernet Consortium (UEC) has defined a new high-performance transport layer to deliver the required performance at scale. A core component of this new standard is the Network Signal-based Congestion Control (NSCC) algorithm. This paper presents SMaRTT, the algorithm that forms the basis of the UEC NSCC specification. SMaRTT is a sender-based congestion control algorithm that systematically combines delay, Explicit Congestion Notification (ECN), and optional packet trimming into a cohesive state machine for fast, fair and precise window adjustments with seamless multipath support. At its core lies the novel QuickAdapt algorithm that accurately estimates and rapidly adapts to available capacity. Our evaluation shows that SMaRTT outperforms existing datacenter congestion control algorithms like Swift, RoCE, and MPRDMA by up to 50\% and provides superior fairness, validating the design choices made in the UEC standard.
\end{abstract}
\section{Introduction}
In the quickly-evolving landscape of AI-centric datacenters, the demand for large-scale and high-performance computing (HPC) capabilities has risen to unprecedented heights. This paradigm shift is demonstrated by the exponential growth of large-scale AI training~\cite{thompson2022computational,oaiscale} and the proliferation of HPC offerings in the cloud~\cite{cloudnoise}. The demand for high throughput at an affordable cost and low latency has become extremely important, and a crucial component to achieving those targets lies in network infrastructure and protocols.




A notable manifestation of this demand is found in the statistics: a staggering 70\% of Azure's traffic is carried by Remote Direct Memory Access (RDMA) technology~\cite{286500}, and all major cloud providers are relying on similar technologies~\cite{9167399,278358,cloudnoise}. Yet, as we witness this transformation, it is worth acknowledging that existing protocols may not always suit the demands of large-scale high-bandwidth networking. For example, \textit{RDMA over Converged Ethernet} (RoCE) is affected by issues like excessive switch buffer requirements for \textit{Priority Flow Control} (PFC), PFC storms, and in-order delivery requirements~\cite{10154243,10.1145/3230543.3230557}. In response to this situation, major tech players have proposed a new transport layer based on Ethernet, called Ultra Ethernet \cite{ultra, hoefler2025ultraethernetsdesignprinciples}. The Ultra Ethernet Consortium has been working to standardize extensions aimed at supporting very large scale (\textgreater{}100K GPUs) high performance AI and HPC datacenter networks. One of the key components include a transport protocol that can support out-of-order data placement, allowing packet spraying to be used to load-balance traffic from a single flow across hundreds of parallel paths, and a standard way of doing packet trimming \cite{179729, 10.1145/3098822.3098825} so that the transport protocol can rapidly detect, NACK and retransmit missing packets while bounding the queue sizes that can build in the network.

\begin{figure}[!t]
  \centering
  \subfloat[2 MiB flows\label{fig:posterchild:a}]{
    \includegraphics[width=0.47\linewidth]{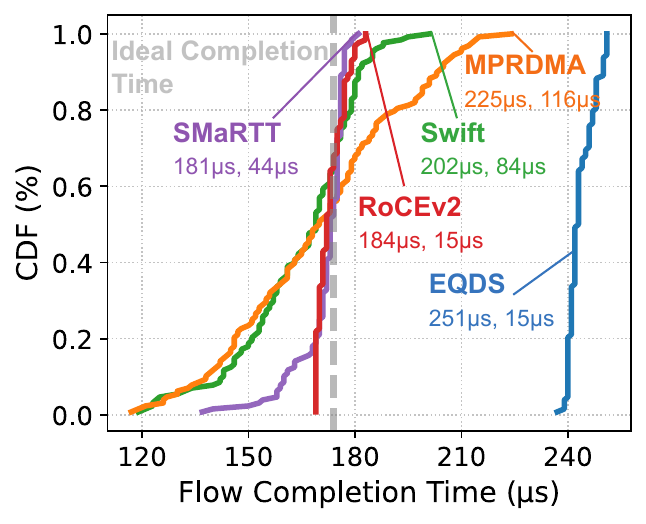}
  }\hfill
  \subfloat[32 MiB flows\label{fig:posterchild:b}]{
    \includegraphics[width=0.47\linewidth]{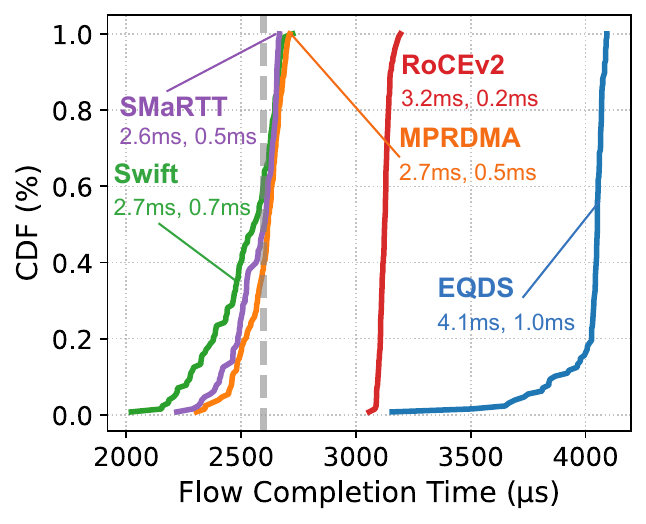}
  }
  \caption{Comparison of SMaRTT with recent CC algorithms on an 8:1 oversubscribed fat tree running a permutation. The first number under each algorithm indicates the overall completion time, while the second is the time difference between the fastest and slowest flow.}
  \label{fig:posterchild}
\end{figure}

This paper details the design and evaluation of SMaRTT, an algorithm that inspired the implementation of Network Signal Congestion Control (NSCC), the standard sender-based congestion control algorithm defined in the Ultra Ethernet 1.0 specifications~\cite{ultra,hoefler2025ultraethernetsdesignprinciples}. We provide the crucial design rationale and empirical validation for the standard's sender-based congestion control.   

AI traffic is notably more bursty and has higher load requirements compared to traditional datacenter traffic. Moreover, when parallelizing AI workloads using collectives such as all-reduces, all the nodes need to wait for the last node to finish before the computation can continue, so it is vital that the tail flow completion times (FCT) are minimized. To address these challenges, we designed the SMaRTT congestion control algorithm to work well with transport protocols that take advantage of multipathing and packet-level adaptive spraying to fully utilize the available network capacity by mitigating the effects of ECMP collisions \cite{facebook}. Although AI workloads can generally tolerate out-of-order delivery, this complicates loss detection. To address this, we propose integrating packet trimming for efficient loss detection and to prevent over-reaction to packet drops \cite{10.1145/3098822.3098825}.

Given the requirements, we build and provide an in-depth analysis of SMaRTT's features, illustrating how each component addresses key challenges: QuickAdapt enables rapid responses to severe congestion, FastIncrease promptly restores the sending rate once congestion subsides, and steady-state management combines ECN and delay signals through three core mechanisms to maintain optimal performance.

On top of that we tightly integrate active load balancing into SMaRTT by designing a transport protocol that deeply integrates congestion control and load balancing. We do so by using a single congestion window that keeps track of the state of all paths.

The Ultra Ethernet simulator source code that contains all modifications described in this paper is publicly available ensuring full reproducibility at \url{https://github.com/ultraethernet/uet-htsim}.

\subsection{Motivation}
We identify the following requirements for a modern high-performance congestion control (CC) algorithm and use them as a starting point to design SMaRTT.
\subsubsection{Fabric Congestion Management}\label{sec:intro:fcm} \sloppy Many CC algorithms have been optimized for managing congestion at the final hop caused by incast~\cite{10.1145/3098822.3098825,278348} due to its relevance for traditional TCP-based workloads~\cite{10.1145/1592681.1592693}. However, with the emergence of new workloads, those algorithms provide sub-optimal performance, since they might have limited visibility of congestion happening in the network fabric. This type of congestion is often the result of oversubscription (i.e., traffic exceeding the capacity of the network), common at upper tiers of datacenter networks~\cite{286425,278358}, but can also occur in a non-oversubscribed setting as a result of ECMP collisions or asymmetries~\cite{conga,drill,10.1145/2535372.2535375}. 
This is shown in Fig.~\ref{fig:posterchild:b}, where we report the CDF of the FCT of 32MiB flows in a permutation scenario with 1,024 nodes, for an 8:1 oversubscribed network. In this scenario, flows are big enough to allow Swift \cite{49448}, a delay based algorithm, and MPRDMA \cite{10.5555/3307441.3307472}, a per-packet ECN based mechanism, to react effectively to congestion. However, when using EQDS \cite{278348} (which uses a NDP-derived \cite{10.1145/3098822.3098825} control loop), FCT is higher due to its inability to adapt to fabric congestion, hence underutilizing network bandwidth due to trimmed packets and re-transmissions.

\subsubsection{Fairness} In this paper, we refer to the term \textit{fairness} as the ability of a CC algorithm to allocate the available bandwidth evenly between competing flows. 
Minimizing jitter and keeping a low and consistent flow completion time (FCT) is particularly important in the context of AI and HPC workloads that exhibit tightly coordinated and synchronized communication patterns. In such systems, performance is often dictated by the slowest node in the system~\cite{netnoise1,netnoise2,tailatscale}, as \textit{"straggler"} flows and growing variability in FCT can have a disproportionately high impact on overall job progress, especially if the congestion control scheme is not able to reclaim bandwidth rapidly when an unfair situation ends.

This is exacerbated by the presence of bursty traffic, which is often common in datacenters. For example, 80\% of RPCs in Google datacenters fit in 1 \textit{bandwidth-delay product} (BDP), and 89\% fit in 4 BDP~\cite{286425}. This behavior is also common in some AI workloads where nodes need to frequently exchange few KiB of data~\cite{10.5555/3571885.3571899}. Algorithms relying on delay to detect congestion often react too late to such transient bursts~\cite{286425,10.1145/3131365.3131375}, leading to unfair treatment of concurrently running flows and increasing completion time. 

We show this effect in Fig.~\ref{fig:posterchild:a}, where we report the \textit{Cumulative Distribution Function} (CDF) of the \textit{Flow Completion Time} (FCT) of 2MiB flows in a permutation scenario with 1,024 nodes, for a 8:1 oversubscribed network (more details on the setup in Sec.~\ref{sec:results}). Using Swift the FCT of the slowest flow is 65\% longer than the fastest flow. MPRDMA suffers from its inherent unfairness which is more visible for small messages. Last, as discussed in Sec.~\ref{sec:intro:fcm}, vanilla EQDS, although more fair, does not manage fabric congestion as effectively as sender-based CC algorithms. On the other hand, SMaRTT quickly and precisely reacts to congestion by relying on a multitude of congestion signals (delay, ECN markings, and packet trimming), improving fairness and overall performance.
   
\subsubsection{Ease of Deployment}
Finally, the rapid scaling of AI infrastructure leads to larger deployments with more endpoints and active flows per endpoint~\cite{10.1145/2829988.2787472}, which can place hard restrictions on the per-flow state that can be stored and manipulated by a congestion control algorithm, especially when implementations target NICs that typically have limited memory resources. This is even more critical when we consider that network bandwidth is rapidly reaching the Tbit/s range and that, for performance reasons, (part of) the CC algorithm would be likely implemented in hardware or firmware directly in the \textit{Network Interface Cards} (NICs). This limits the amount of memory the algorithm can use and the complexity of the algorithms that can be executed (at 800Gbit/s, with 4KiB packets, the NIC needs to process one packet every 40 nanoseconds).

\sloppy
Additionally, gradually deploying new switch generations or incorporating switches from diverse vendors with varying features is common in datacenters~\cite{10.1145/3464994.3464996}. Consequently, CC algorithms relying on specific new features demanding switch hardware support~\cite{10.1145/3341302.3342085,10.1145/3386367.3431316,276958} might have difficulties getting adopted. 

\subsection{Contributions}
To fulfill these requirements, we introduce a new CC algorithm we call SMaRTT (\textit{\underline{S}ender-based \underline{Ma}rked \underline{R}apidly-adapting \underline{T}rimmed \& \underline{T}imed Transport}). SMaRTT is a sender-based CC algorithm designed to run over lossy Ethernet. To ease deployment and allow for seamless operation in heterogeneous environments that potentially include different generations of network equipment, SMaRTT does not require any advanced switch capabilities, such as in-band network telemetry (INT) \cite{10.1145/3341302.3342085}. It instead leverages a combination of ECN-marking and round-trip-time (RTT) measurements, two signals that have been commonly used in a mutually exclusive manner despite each of them possessing distinct valuable benefits that complement one another (as we show in Sec.~\ref{smart:ccsignals}). Last, SMaRTT efficiently uses memory and compute resources, since it only needs to store a few bytes per flow and does not require any \textit{pull queue} or specific control packets, as common in many receiver-driven algorithms. Specifically, SMaRTT makes the following key contributions:

\paragraph{QuickAdapt} A novel mechanism that can quickly converge to the achievable bandwidth capacity at any bottleneck, typically within one RTT of receiving a severe congestion signal (Sec.~\ref{sec:quick_adapt_text}). QuickAdapt relies on packet trimming or, in the case where it is not supported, on delay. Using QuickAdapt, SMaRTT can react to severe congestion as quickly as a receiver-based protocol but without requiring any special support on the receiver side (e.g., pull queues). We also argue that this feature can be retrofitted to any transport mechanism operating within tight RTT bounds, thus significantly reducing tail latencies. In particular, QuickAdapt can reduce queue occupancy much earlier than other sender-based mechanisms and hence packet drops.

\paragraph{Fair Increase} A new congestion window management technique to improve bandwidth fairness (Sec.~\ref{sec:fair_increase}). In a nutshell, the congestion window is increased or decreased more conservatively (and proportionally to the currently acknowledged packet size) when the ECN and delay congestion signal disagree on the presence of congestion.

\paragraph{Tightly Coupled Load Balancer} The design of SMaRTT tightly integrates congestion control and load balancing by using a single congestion window for all paths of a connection. This contrasts with approaches like MPTCP~\cite{mptcp}, which maintain separate congestion windows for each path. Effectively, in SMaRTT, the congestion window tracks the available capacity in the network across the aggregate of paths. On top of that we use the same congestion signal (ECN) to drive the load balancer mechanism. Finally, SMaRTT always tries to let the load balancer find a new path when congestion is starting to build up before actually starting to reduce its window. This enables SMaRTT to tolerate transient queuing resulting from ECMP hash collisions or imperfect load balancing, engaging only when load balancing alone is insufficient to mitigate congestion.

\paragraph{Fast Increase} A new mechanism that can quickly increase the congestion window when the sender detects that the network is congestion free by looking at both the recent delay and ECN signals (Sec.~\ref{sec:fast_increase}). This has two major benefits: 1) in case where uneven incasts happen, some flow will finish later than others. With Fast Increase, such flows would quickly recover their sending rate 2) while SMaRTT tries to keep fairness between flows, sometimes one flow could still slightly lag behind. With Fast Increase, we make sure to limit the impact of such unfairness.

\section{Background}
The behavior of CC algorithms is primarily determined by how they detect (Sec.~\ref{sec:back:signals}) and react (Sec.~\ref{sec:back:control}) to congestion.

\subsection{Congestion Signals}\label{sec:back:signals}
\paragraph{Delay} End-to-end delay accurately approximates congestion. Delay can be computed either by the sender, measuring the RTT (e.g., in Swift~\cite{49448} and TCP~\cite{1092259}), or by the receiver, annotating acknowledgment packets (ACK) with one-way delay information (e.g., in TIMELY~\cite{43840} or DX~\cite{7544640}). \tom{In the following, we assume that accurate delay measurements are obtained via NIC timestamping~\cite{49448,286500}. Since SMaRTT derives RTT measurements from relative differences between timestamps taken by the same NIC, precise synchronization across different NICs is not strictly necessary.}

\paragraph{ECN-marking} With \textit{explicit congestion notification} (ECN), switches set a bit in the traffic class field of the IP header when a packet experiences congestion. The receiver relays this information, as a flag, back to the sender, which adjusts its rate accordingly~\cite{zhu2015congestion,10.1145/1851275.1851192,8847013}. Because ECN notifies congestion using a single bit, it provides less information than time-based signals, which instead can approximate the queuing delay along the path. Switches can use different policies to decide if a packet must be marked. For example, in \textit{random early detection} (RED)~\cite{251892}, switches randomly mark packets with a probability linear in the switch queue size, if that size is within two thresholds ($K_{min}$ and $K_{max}$). Although ECN was designed to mark packets when they are enqueued~\cite{rfc3168}, it has been shown that doing that when dequeued allows CC algorithms to react faster~\cite{Zhu2016ECNOD}. Dequeue marking can be easily implemented on most existing switches~\cite{wu2012tuning}, and in the rest of the paper, we assume that switches use RED and mark ECN packets when dequeued. Although ECN-based CC algorithms can react quicker than delay-based ones~\cite{Zhu2016ECNOD}, they perform poorly when dealing with incast and can be challenging to tune~\cite{49448}. 

\paragraph{Packet Losses} Packet losses have extensively been used to detect severe congestion~\cite{251892,10.1145/2208917.2209336}. \tom{However, relying solely on packet losses causes the algorithm to react too late to congestion. Losses are also typically detected via timeouts, which are difficult to tune and may lead to unnecessary retransmissions. Additionally, in a multi-path sprayed environment, tracking received PSNs at the receiver is insufficient for reliably detecting packet loss.}

\paragraph{Packet Trimming} With packet trimming, a switch can remove parts of the packet (e.g., payload) rather than dropping it. Packet trimming retains essential information (e.g., headers), allowing the host to quickly detect and react to congestion~\cite{179729,10.1145/3098822.3098825,278348}. Past research has demonstrated that packet trimming can be enabled on switches such as Intel Tofino, the Broadcom Trident 4 and NVidia Spectrum 2 by leveraging their ability to reroute packets to an alternative port when the initial egress queue becomes saturated~\cite{adrian2022implementing}.


\subsection{Rate Control}\label{sec:back:control}
The decision on how to react to congestion can be made either by the receiver~\cite{10.1145/3098822.3098825,278348,10.1145/2716281.2836086,9047463,10.1145/2807591.2807600,10.1145/3230543.3230564}, the sender~\cite{49448,zhu2015congestion,10.1145/1851275.1851192,9796803} or the switch~\cite{10.1145/3386367.3431316,276958}. Whereas receiver-based algorithms can precisely regulate the transmission of the different senders (e.g., in incast scenarios), they require extra control packets and data structures (e.g., pull queues) on the receiver side. Also, some of these algorithms can fall short when dealing with fabric congestion~\cite{49448}. On the other hand, sender-based schemes do not require such extra complexity but cannot precisely regulate the transmission rate during incast traffic due to lack of interaction with the other senders.

As for rate enforcement, the sender can either change the transmission rate directly or modify its congestion window (cwnd) which sets a limit on the inflight bytes.

SMaRTT is a sender-based and window-based congestion control mechanism.


\section{SMaRTT Design}
\label{sec:smartt-design}
\paragraph{Overview} SMaRTT is a high-performance, easy-to-deploy, and lightweight CC algorithm that can quickly react to congestion happening both at the last hop (incast) and in any other part of the network. SMaRTT keeps a number of in-flight bytes equal to the size of a \textit{congestion window} (\textit{cwnd}). SMaRTT updates the window depending on the estimated network congestion, striving to keep RTT lower than a pre-determined and constant target RTT. This can be expressed relatively to the \textit{base RTT} (i.e., the RTT under idle network conditions). The base RTT can either be measured during datacenter setup (once per each path hop count) or dynamically updated, similar to Swift~\cite{49448}. Throughout the rest of the paper, we assume a target RTT equal to $1.5$x the base RTT.

\paragraph{Required Features} SMaRTT relies on widely available features like ECMP, ECN marking, and packet trimming. ECMP and ECN marking are available on all contemporary datacenter switches, e.g., ~\cite{bcomecmp,spectrum,10154243}, whereas packet trimming can be implemented on some existing switches~\cite{adrian2022implementing} and is starting to be natively supported by the latest generation of switches \cite{trimming_rdma}. If ECMP is not supported, we can still run the CC algorithm. If packet trimming is not supported, SMaRTT falls back to timeouts, with limited performance impact (Sec.~\ref{sec:trimming}). We note that alternative loss-detection strategies beyond timeouts could be used to further improve performance without relying on trimming, but these are orthogonal to the focus of this paper. Similar to EQDS~\cite{278348}, we assume that each switch port has a high-priority queue to forward control packets such as trims and ACKs, and a lower-priority queue to forward data packets. By doing so, packets carrying potential congestion information are prioritized so that SMaRTT can react quickly. We design SMaRTT to work with packet spraying-based load balancers such as Random Packet Spraying \cite{6567015} or adaptive versions such as REPS \cite{bonato2025repsrecycledentropypacket}.

\paragraph{Assumptions} This work mainly focuses on widely used fat tree networks~\cite{51587,cloudnoise}. However, SMaRTT could be adapted to work with other topologies such as Torus, Dragonfly~\cite{4556717,278358,9355230}, BCube~\cite{10.1145/1592568.1592577}, SlimFly~\cite{10.1109/SC.2014.34}, HammingMesh~\cite{10.5555/3571885.3571899} and others. Moreover, we intentionally limit buffering to a worst-case scenario, assuming that each per-port queue is one \textit{bandwidth-delay product} (BDP) in size. 

Finally, this work focuses on lossy networks since that is the expected main operating mode in the UE protocol \cite{ultra}. However, in future work, SMaRTT could be extended to operate in lossless networks with PFC by relying solely on ECN and trimming as congestion signals.

\begin{figure}[ht!]
        \begin{center}
            \includegraphics[width=1\columnwidth]{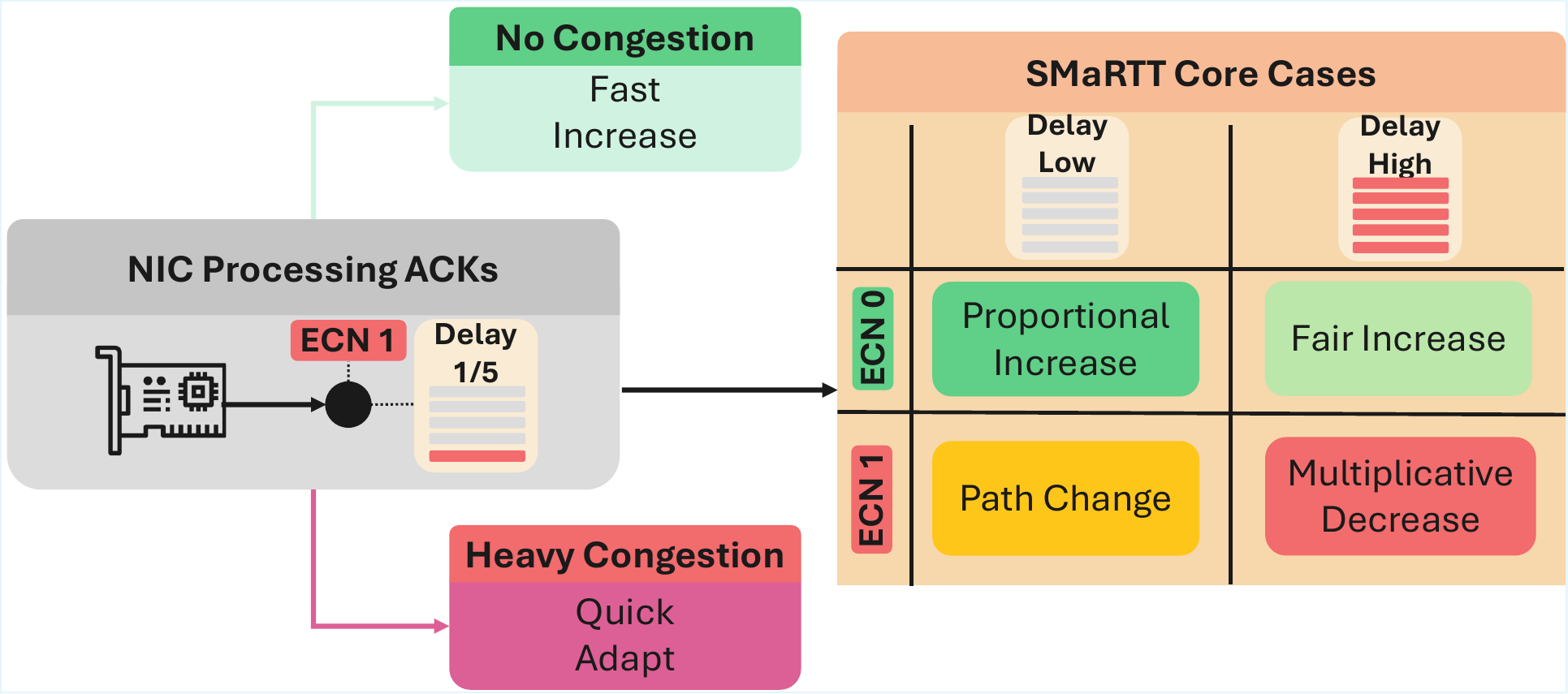}
        \end{center}
        \caption{High level block diagram of SMaRTT.} 
        \label{fig:smartt_scheme}
        \vspace{-1.2mm}
\end{figure}

 \paragraph{Presentation} In the following, we motivate the choice of using spraying (Sec.~\ref{sec:lb}) combined with both delay and ECN as congestion signals (Sec.~\ref{smart:ccsignals}), and then describe each combination of them in the main control loop (Sec.~\ref{sec:mainloop}). We then present \textit{QuickAdapt} (Sec.~\ref{sec:quick_adapt_text}) and \textit{FastIncrease} (Sec.~\ref{sec:fast_increase}), two techniques we introduce to further improve SMaRTT reactiveness. Eventually, we discuss how to select SMaRTT parameters (Sec.~\ref{sec:tuning}).
Figure~\ref{fig:smartt_scheme} includes a high-level block diagram of SMaRTT input signals and reaction options.

\subsection{Load Balancing Interaction} \label{sec:lb}
\begin{figure}[h]
    \centering
    \includegraphics[width=0.80\linewidth]{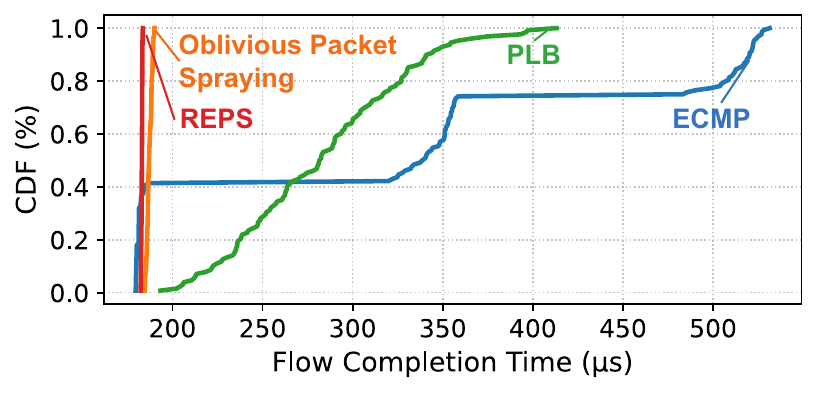}
    \caption{Permutation of SMaRTT for different load balancers.}
    \label{fig:ablation:multipathing}
\end{figure}

Load balancing plays a crucial role in mitigating network hot spots and congestion, even before congestion control mechanisms come into effect. For SMaRTT, we decide to pair it with packet spraying based approaches \cite{6567015} such as Recycled Entropy Packet Spraying (REPS) \cite{bonato2025repsrecycledentropypacket} to fully utilize the multipathing capabilities of the network. This is because single path algorithms such as ECMP \cite{ecmp} or even more advanced variants like PLB \cite{plb} struggle because of collisions to fully utilize the network. This is further motivated in Figure~\ref{fig:ablation:multipathing}, which shows the different performance during a permutation workload. REPS essentially works by re-using good paths that arrive back at the sender without ECN mark while re-routing new packets when a certain path arrives back at the sender with the ECN bit set. We choose REPS since the 1.0 Ultra Ethernet specification explicitly cites it as a reference load-balancing mechanism for Ultra Ethernet Transport (UET)~\cite{ultra,hoefler2025ultraethernetsdesignprinciples}.

SMaRTT, as we can see in Figure~\ref{fig:smartt_scheme} and with more details in Section~\ref{sec:lb_change}, will not change its congestion window when it initially receives packets with ECN but with low delay. This is to give the adaptive load balancer, such as REPS, a possibility to change path. \tom{However, if the delay continues to rise and exceeds the target RTT threshold, SMaRTT responds with a multiplicative decrease in the congestion window and may additionally invoke QuickAdapt in cases of severe congestion.}

\tom{While this paper does not focus on load balancing, we note that SMaRTT is designed to be compatible with other load balancers, ideally those that operate on a per-packet basis and can adaptively reroute around congested or failed paths.}

\subsection{SMaRTT Congestion Signals}\label{smart:ccsignals}
Choosing the congestion signals to react to is critical for the design of a congestion control protocol. SMaRTT uses \textit{both} ECN and delay (RTT), and it can also leverage packet trimming to react to severe congestion, which we discuss in Sec.~\ref{sec:trimming}. 

\textit{Why use both signals?} \tom{As noted in prior work~\cite{Zhu2016ECNOD}, ECN offers several advantages. When applied at \textit{packet dequeue}, it provides more up-to-date feedback about queue conditions than delay-based signals. Moreover, ECN is less sensitive to transient network slowdowns: while the signal itself may be delayed, its meaning remains unaffected, unlike delay signals, which can be distorted by unrelated latency. Finally, delay can appear elevated due to transient packet collisions or queuing delays that accumulate across multiple hops, even when no single hop is significantly congested, making it a less reliable congestion indicator.} 

\begin{figure}[htpb]
        \begin{center}
            \includegraphics[width=0.95\columnwidth]{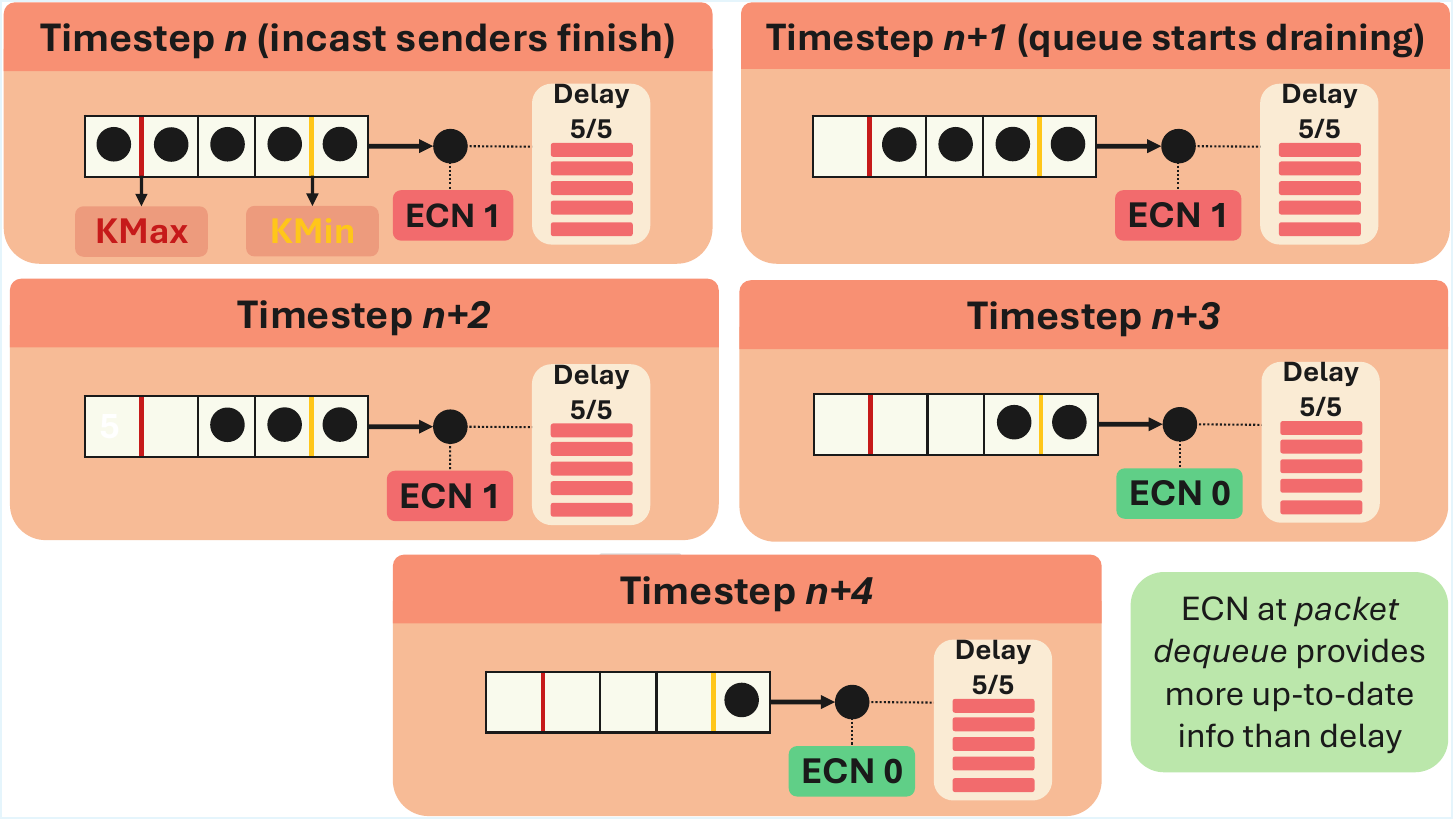}
        \end{center}
        \caption{Comparing ECN on egress and delay-based congestion signals during an incast.} 
        \label{img:ecnrtt}
        \vspace{-1.2mm}
\end{figure}

\tom{In Figure~\ref{img:ecnrtt}, we showcase with a simple example how ECN (applied at packet dequeue) can return more up-to-date queue information than delay. In particular, we simulate an incast scenario that has been active long enough to fill the queue and has just completed, initiating the queue’s draining phase. As shown, at \textit{Timestep n}, the dequeued packet correctly carries both an ECN mark and a maxed out delay value, reflecting that the queue had been full for some time. However, from \textit{Timestep n+3} onward, ECN and delay begin to diverge. This occurs because ECN reflects the queue state at the moment of dequeue, providing timely feedback, while delay remains elevated until the last packet exits the previously full queue, even if no congestion is present at dequeue.}

However, ECN is not without its own limitations. Specifically, the ECN signal provides only a single-bit indication, which is far less granular than the multi-bit approach offered by delay-based signals. On top of that, at small congestion window values, ECN gets very fragmented feedback which is also heavily influenced by its probabilistic behaviour, making it susceptible to sub-optimal behaviour at small congestion windows. Finally, other works have noted how ECN can start to struggle once topology grow to have multiple tiers (usually above 3) \cite{Zeng2017CombiningEA}.

\begin{figure}[!t]
  \centering
  \includegraphics[width=\linewidth]{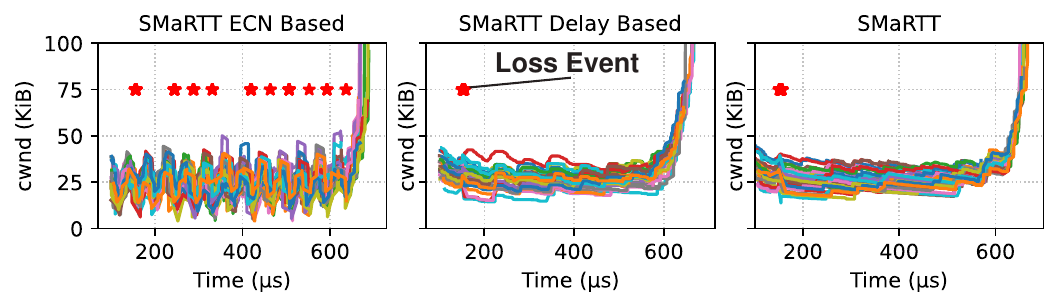}
  \caption{\textit{cwnd} evolution over time for three variants of SMaRTT. The red stars indicate when NACKs happen.}
  \label{fig:ecnvsrtt}
\end{figure}

We then proceed to show how with a large incast, the ECN signal can start to struggle. This is because, as previously noted, when the congestion window gets very small (in our large incast, the ideal value is around 5 packets), its signal becomes fragmented and unreliable. In Figure~\ref{fig:ecnvsrtt} we show this by comparing three versions of SMaRTT. We note that we show only the plot after initial convergence to show how ECN is: 1) more unstable during steady state at low \textit{cwnd} 2) can cause packet drops even during steady state. On the other hand, the delay version of SMaRTT and the full SMaRTT implementation are stable during steady state with the combined version doing slightly better.

For these reasons we decide to combine both signals in SMaRTT. We take as baseline a per-packet approach that uses ECN similar to MPRDMA \cite{10.5555/3307441.3307472} and we enhance it with RTT. To summarize we derive four cases to combine ECN and delay where we look at all the possible combinations of them. This creates our congestion control core loop which is summarized at a high level with Algorithm~\ref{alg:core} and Figure~\ref{fig:smartt_scheme}.
\begin{algorithm}
\caption{\small SMaRTT Core Cases}
\begin{algorithmic}
\footnotesize
\captionsetup{font=small}
\Procedure{core\_cases}{p}
    \If{$p.ecn$ \textbf{and} $p.rtt > trtt$}
        \State \textsc{multiplicative\_decrease}()
        \Comment{\textit{Sec.~\ref{sec:multiplicative_decrease}. Only once per base RTT.}}
    \ElsIf{$p.ecn$ \textbf{and} $p.rtt < trtt$}
        \State \textsc{load\_balancer\_path\_change}()
        \State \textsc{no\_op\_cc}()
    \ElsIf{\textbf{not} $p.ecn$ \textbf{and} $p.rtt > trtt$}
        \State \textsc{fair\_increase}()
        \Comment{\textit{Sec.~\ref{sec:fair_increase}}}
    \ElsIf{\textbf{not} $p.ecn$ \textbf{and} $p.rtt < trtt$}
        \State \textsc{proportional\_increase}()
        \Comment{\textit{Sec.~\ref{sec:multiplicative_increase}}}
    \EndIf
\EndProcedure
\end{algorithmic}
\label{alg:core}
\end{algorithm}

\subsection{Fair Increase}
After having established the four core cases of SMaRTT, we analyze in detail some aspects of it. 

In Fair Increase the ACK is not ECN-marked, but reports an RTT higher than the target. This scenario represents a critical state of signal ambiguity that purely delay-based or ECN-based protocols handle poorly. The high RTT indicates persistent queuing exists somewhere in the aggregate of paths, while the non-ECN-marked ACK provides fresh, path-specific evidence that the queue at the bottleneck for this specific packet has recently drained.

A naive delay-based protocol might incorrectly initiate a multiplicative decrease based on the stale, high RTT, throttling a potentially clear path and reducing throughput. Conversely, ignoring the high RTT and performing a large increase would be reckless, risking immediate re-congestion.

SMaRTT's Fair Increase resolves this ambiguity by acting as a conservative probing mechanism. It trusts the most recent signal (the non-ECN mark) enough to avoid a rate decrease, but respects the historical warning from the high RTT by increasing the window minimally. This allows the sender to cautiously test for available capacity on the newly cleared path without destabilizing the system, ensuring both stability and fairness, especially in dynamic, multi-path environments where RTT measurements can lag behind real-time path conditions. This prevents oscillations and maintains higher average throughput compared to a protocol that would over-react with a multiplicative decrease.

\begin{figure}[!t]
  \centering

  \subfloat[Fair increase\label{fig:ablation:fi-vs-ai:fi}]{
    \begin{minipage}{0.47\linewidth}
      \centering
      \includegraphics[width=\linewidth]{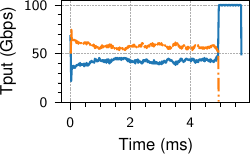}
      \vspace{0.6ex}
      \includegraphics[width=\linewidth]{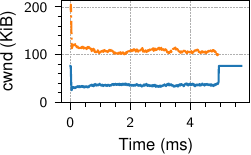}
    \end{minipage}
  }\hfill
  \subfloat[Additive increase\label{fig:ablation:fi-vs-ai:ai}]{
    \begin{minipage}{0.47\linewidth}
      \centering
      \includegraphics[width=\linewidth]{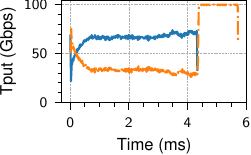}
      \vspace{0.6ex}
      \includegraphics[width=\linewidth]{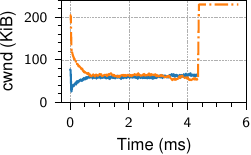}
    \end{minipage}
  }

  \caption{We get better throughput fairness with fair increase vs.\ additive increase. The blue flow has $3\times$ lower base RTT than the orange flow.}
  \label{fig:ablation:fi-vs-ai}
\end{figure}

On top of that, this feature helps reaching fairness when flows have different RTTs (for instance in the case of local and remote flows competing for the same bottleneck). The goal is to help flows stabilize faster and ensure stability (more details in Sec.\ref{sec:mainloop}).
To showcase this aspect we consider the case when flows with different base RTTs compete for a bottleneck path. As shown in Fig.~\ref{fig:ablation:fi-vs-ai}, with additive increase, the flows converge to having the same \textit{cwnd} and throughput \textit{(cwnd/RTT)} inversely proportional to the RTTs, whereas with fair increase, the flows converge to \textit{cwnds} that are proportional to their base RTTs, and so the throughput is roughly similar.

\subsection{Packet Trimming} \label{sec:trimming}
\begin{figure}
    \centering
    \includegraphics[width=0.85\linewidth]{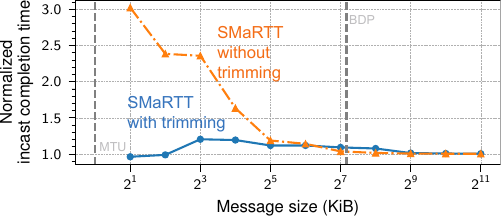}
    \caption{With trimming, the incast completion times are shorter by roughly a constant amount (RTO).}
    \label{fig:ablation:trim-vs-timeout}
    \vspace{-1.2mm}
\end{figure}

We shift our attention to trimming, one of the key components of SMaRTT. When using packet trimming, the switch removes the payload and forwards the header if the queue size exceeds a threshold~\cite{10.1145/3098822.3098825}. This allows congestion information to be communicated to the receiver even on congested networks. With SMaRTT, we trim packets only if they would otherwise be dropped because of a full buffer. Trimming helps SMaRTT deliver faster feedback about losses, which, as a consequence, allows for faster convergence and less packet losses. If we do not have trimming enabled and rely on timeouts, the CCA learns of the packet loss after the timeout period (roughly 84 us = 7 base RTTs, in the worst case 6 BDPs of queues build up and 1 BDP of in-flight packets giving a total of maximum 7 BDPs), whereas with trimming, the CCA learns of the packet loss roughly half to one base RTT after the loss happens (as the trimmed packets have higher priority, they don't wait behind queues). As a result, the CCA can react 6 base RTTs sooner. As shown in Fig.~\ref{fig:ablation:trim-vs-timeout}, this is consistent with our observation that with timeouts, the completion times for flows are roughly 6 base RTTs higher when there are situations with loss (heavy congestion), e.g., incast. \tom{However, this difference becomes negligible for large message sizes, as the timeout duration becomes small compared to the overall incast runtime. Finally, trimming further tightens the congestion feedback loop: unlike timeouts, which make it unclear whether the congestion control has already reacted or still needs to, trimming provides more immediate and unambiguous signals.}

\subsection{QuickAdapt} \label{sec:quick_adapt_text}
SMaRTT reacts to heavy congestion by quickly adjusting the congestion window using a novel technique, which we call \textit{QuickAdapt}. QuickAdapt behaves similarly to receiver-based CC algorithms without needing special support on the receiver side (e.g., pull queues) or per-packet states like in BBR~\cite{45814}. The key idea behind it is that in case of heavy incast congestion, we do not want the sender to slowly reduce its sending window until convergence as this could result in a massive amount of dropped packets and re-transmissions, even when using packet trimming.

\begin{figure}
    \centering
    \includegraphics[width=0.85\linewidth]{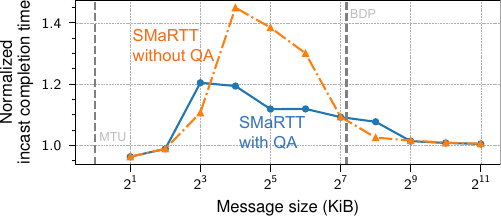}
    \caption{QuickAdapt helps SMaRTT to converge quickly during large incasts. This helps curb queue buildup and losses, yielding improvement in FCTs.}
    \label{fig:ablation:qa-vs-noqa}
    \vspace{-1.2mm}
\end{figure}

QuickAdapt maintains a running count of the bytes that have been received over the last $\mathit{trtt}$ window and, if a loss is detected with a trim packet, it sets the congestion window to that value after finishing its current measurement. It is possible to scale that value by a constant ($\mathit{qa\_scaling}$) depending on the desired queue occupancy (see Sec.~\ref{sec:tuning}). In our case, we set the scaling value to be $1$ as this implies that the windows will be dropped to a value that is directly related to our $trtt$. In fact, if we were to measure for only one $brtt$ before adjusting the window then we would drop our congestion windows to a value which is roughly equivalent to running the wire at line rate but with no queuing. However, this could result in under-utilization since there is some noise in our measurement. For this reason, we measure for longer (one $trtt$ which is logically equivalent to dropping the windows to keep the queues at half of their sizes). Figure~\ref{fig:ablation:qa-vs-noqa} shows the overall benefit provided by QuickAdapt. 

\begin{figure}[!t]
  \centering
  \includegraphics[width=\linewidth]{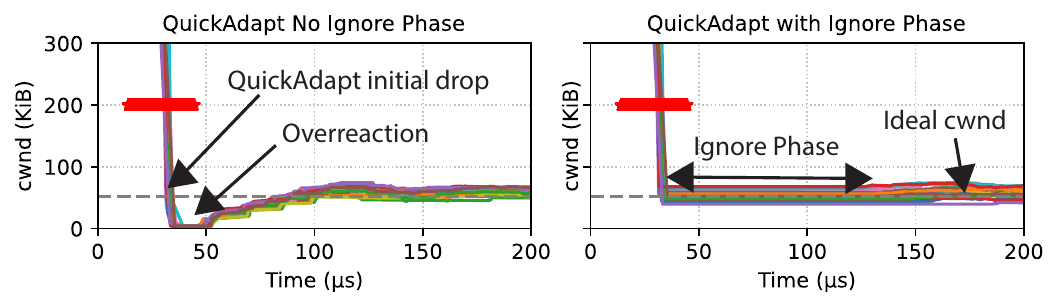}
  \caption{\textit{cwnd} evolution over time for an incast with QuickAdapt Ignore off and on. The dotted line indicates the ideal \textit{cwnd} value. The red dots indicate loss events.}
  \label{fig:qa_ignore}
\end{figure}

When SMaRTT triggers a QuickAdapt adjustment, the window can be drastically reduced (i.e., in the case of a 16:1 incast, the window will be, after only one $trtt$ of receiving the first packet, dropped to $1/16$ of its original value. While this happens, the network will likely still be full of packets returning with ECN marks and high delay. However, it would not make sense to further reduce the congestion window as QuickAdapt had already dropped it to the correct value. For this reason, when we first trigger QuickAdapt, we set a simple counter to ignore all the bytes that are in-flight when it comes to congestion control information. By doing so, QuickAdapt does not cause SMaRTT to overreact to congestion and, for the same reason, it is applied at most once per target RTT. To showcase the issue, we present in Figure~\ref{fig:qa_ignore}, how congestion windows are quickly dropped thanks to QuickAdapt but then showcase the difference with and without the ignore phase. The QuickAdapt pseudocode is presented in Algorithm~\ref{alg:qa_algo}. 

\begin{algorithm}[htpb]
\footnotesize
    \captionsetup{font=small}
    \caption{\small QuickAdapt Pseudocode}
    \begin{algorithmic}[1]
        \Procedure{quick\_adapt}{p}
            \State \textit{adapted} = \textbf{false}
            \If{$\text{now} \geq \text{end}$} \Comment{\textit{QuickAdapt at most once per target RTT}}
                \If{\textit{trigger\_qa} \textbf{and} \textit{end} $\neq$ {0}}
                    \State \textit{trigger\_qa} = \textbf{false}
                    \State \textit{adapted} = \textbf{true}
                    \State \textit{cwnd} = \textbf{max}(\textit{acked}, \textit{mtu})
                    \State \textit{bytes\_to\_ignore} = \textit{unacked} \Comment{\textit{Ignore next congestion signals}}
                    \State \textit{bytes\_ignored} = {0}                    
                \EndIf
                \State \textit{end} = \textit{now} + \textit{trtt}
                \State \textit{acked} = {0} 
            \EndIf
            \State \Return \textit{adapted}
        \EndProcedure
    \end{algorithmic}
    \label{alg:qa_algo}
\end{algorithm}

\paragraph{QuickAdapt without trimming} \label{sec:eval:trimvstimeout}
If trimming is not supported, SMaRTT falls back to a simple timeout-based approach for losses, requiring minimal changes to the overall logic. However, for QuickAdapt, we cannot rely on timeouts due to congestion loss events as they might happen significantly after the beginning of a heavy congestion event. Instead, in order to trigger QuickAdapt we look at the current RTT and the amount of received bytes in the current time window. If the RTT is high and the amount of $acked$ bytes is low compared to the $cwnd$ value then we trigger it. The ignore phase of QuickAdapt becomes even more important in the case of timeouts as we need to ignore such timeouts after dropping the cwnd to the QuickAdapt value. 
\vspace{-1.2mm}
\subsection{FastIncrease}\label{sec:fast_increase}
\begin{figure}[!t]
  \centering
  \subfloat[With fast increase]{%
    \includegraphics[width=0.48\linewidth]{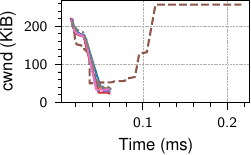}%
  }\hfill
  \subfloat[Without fast increase]{%
    \includegraphics[width=0.48\linewidth]{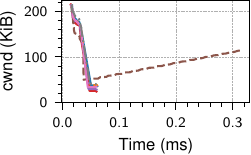}%
  }
  \caption{Fast increase allows faster convergence when flows finish.}
  \label{fig:ablation:fi-vs-nofi}
  \vspace{-1mm}
\end{figure}

\textit{FastIncrease} quickly reclaims available bandwidth after the completion of some flows (e.g., in case of incast with uneven sizes). FastIncrease is triggered when at least $\mathit{cwnd}$ contiguous bytes did not experience any congestion (i.e., the RTT is close to the base RTT, and the ACKs are not ECN marked). In that case, as long as FastIncrease stays active, the congestion window is increased by $k$ MTUs for each ACK. $k$ is a constant that we set equal to $2$ (after tuning, Sec.~\ref{sec:tuning}). We show the FastIncrease pseudocode in Algorithm~\ref{alg:fi}. \tom{In Figure~\ref{fig:ablation:fi-vs-nofi} we show an example of FastIncrease helping in the case of uneven incasts (common in Alltoallv patterns \cite{bursty1,bursty2}). In particular, we have one flow (represented by the dotted line) which is twice as long as the other ones. With fast increase, such flow quickly reclaims the available bandwidth by raising its cwnd.}

\begin{algorithm}[htpb]
\footnotesize
    \captionsetup{font=small}
    \caption{\small FastIncrease Pseudocode}
    \begin{algorithmic}[1]
    \Procedure{fast\_increase}{p}
        \If{$\mathit{rtt} \approx \mathit{brtt}$ \textbf{and not} $\text{p.ecn}$}
            \State $\mathit{count} \mathrel{+}= \mathit{p.size}$
            \If{$\mathit{count} > \mathit{cwnd}$ \textbf{or} $\mathit{increase}$}
                \State $\mathit{cwnd} \mathrel{+}= k \cdot \mathit{mtu}$
                \State $\mathit{increase}$ = \textbf{true}
                \State \Return \textit{increase}
            \EndIf
        \Else
            \State $\mathit{count} = 0$
            \State $\mathit{increase}$ = \textbf{false}
        \EndIf    
        \State \Return \textit{increase}
        \EndProcedure
    \end{algorithmic}
    \label{alg:fi}
\end{algorithm}
\vspace{-1.2mm}

\vspace{-1mm}

\subsection{Adjustment Period}
We design SMaRTT to be a per-packet congestion control algorithm in order to improve its reaction times and avoid making sudden changes frequently. This helps stability as, beside QuickAdapt and FastIncrease, it is not possible to experience large bandwidth changes in SMaRTT. However, reacting on a per-packet basis might be problematic in real hardware, especially when using the latest switches capable of 400Gbps or 800Gbps. For these reasons, we also test versions of SMaRTT where we either accumulate values per packet but update the cwnd less frequently to avoid executing divisions per-packet (more hardware expensive) or we sample every $n$ packets and then scale the response accordingly. We find that both work well in our testing, with SMaRTT working well even when sampling every 8 packets or accumulating for up to an RTT worth of time.

\subsection{Overall Picture} Algorithm \ref{alg:smartt_algo} summarizes the core SMaRTT logic. So far, we discussed all the main features of SMaRTT and left the core control loop to be studied in detail in the next section. It is worth remarking that SMaRTT has a low memory footprint (19 bytes per flow, and 28 bytes globally) to ensure scalability and ease of hardware implementation on NICs. Moreover, to evaluate SMaRTT's computational cost, we compiled SMaRTT's main loop on an Intel Xeon X7550 CPU using \texttt{gcc} 10.2.1. We measured a number of per-packet instructions executed ranging from 59 (when QuickAdapt is triggered) to 94 (when one of the four core cases is executed).

    \begin{algorithm}[htpb]
    \footnotesize
    \captionsetup{font=small}
    \caption{\small SMaRTT Pseudocode}
    \begin{algorithmic}[1]
    \Procedure{congestion\_loop\_logic}{p}
    \State $\mathit{acked} \mathrel{+}= \mathit{p.size}$
    \State $\mathit{bytes\_ignored} \mathrel{+}= \mathit{p.size}$
    \State
    \If{\textit{p.is\_ack}}
        \If{$\mathit{bytes\_ignored} < \mathit{bytes\_to\_ignore}$}
            \State \Return \Comment{\textit{Ignore ACKs after QuickAdapt (Sec.~\ref{sec:quick_adapt_text})}}
        \EndIf
        \State $\mathit{adp} = \text{quick\_adapt(p)}$ \Comment{\textit{Sec.~\ref{sec:quick_adapt_text}}}
        \State $\mathit{finc} = \text{fast\_increase(p)}$ \Comment{\textit{Sec.~\ref{sec:fast_increase}}}
        \If{$\mathit{adp}$ \textbf{or} $\mathit{finc}$}
            \State \Return \Comment{\textit{Exit if QuickAdapt or FastIncrease triggered}}
        \EndIf
        \State $\text{core\_cases(p)}$
        \Comment{\textit{Sec.~\ref{sec:mainloop}}}
    \ElsIf{$\mathit{p.is\_trimmed}$} 
    \Comment{\textit{Sec.~\ref{sec:quick_adapt_text}}}
        \State $\mathit{cwnd} \mathrel{-}= \mathit{p.size}$
        \State \textit{trigger\_qa} = \textbf{true}
        \State $\text{retransmit\_packet(p)}$
        
        \If{$\mathit{bytes\_ignored} >= \mathit{bytes\_to\_ignore}$}
            \State $\text{quick\_adapt(p)}$ \Comment{\textit{Sec.~\ref{sec:quick_adapt_text}}}
        \EndIf        
    \EndIf
    \State \textit{cwnd} = \textbf{max}(\textbf{min}(\textit{cwnd}, \textit{bdp}), \textit{mtu}) \Comment{\textit{Sec.~\ref{sec:tuning}}}
    \EndProcedure
\end{algorithmic}
\label{alg:smartt_algo}
\end{algorithm}
\subsection{Core Cases Deep Dive}\label{sec:mainloop}
The main control loop (\S\ref{sec:smartt-design}) of SMaRTT has four core cases: \textit{Load Balancer Activation} (\S\ref{sec:lb_change}), \textit{Multiplicative Decrease} (\S\ref{sec:multiplicative_decrease}), \textit{Fair Increase} (\S\ref{sec:fair_increase}), \textit{Proportional Increase} (\S\ref{sec:multiplicative_increase}).

\subsubsection{Load Balance Path Change}\label{sec:lb_change} The ACK is \textbf{ECN-marked} and reports an \textbf{RTT smaller than the target one}. This case indicates a growing queue (signaled by ECN) that has not yet grown enough to significantly impact packet delay (as RTT is below target). As a consequence, SMaRTT activates the load balancer, such as REPS, to change paths while keeping the congestion window untouched. This rationale is to first test if the congestion is a transient load-balancing issue before resorting to a \textit{cwnd} reduction.

\begin{figure}[ht]
  \centering
  \subfloat[4MiB Permutation without oversubscription\label{fig:smartt_load:a}]{
    \includegraphics[width=\linewidth]{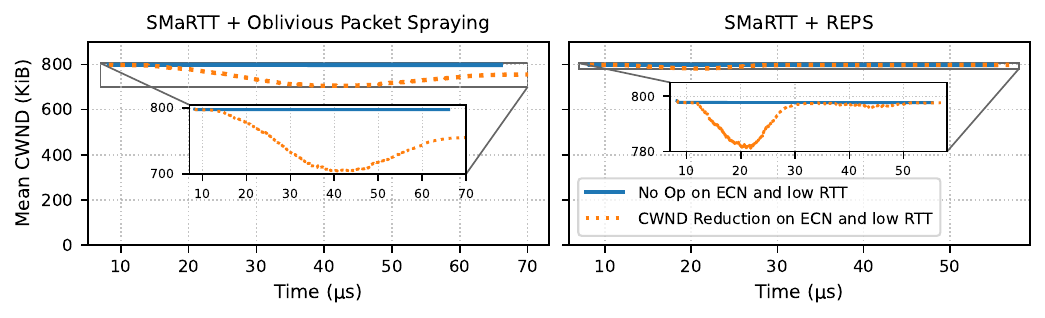}
  }
  \par\medskip
  \subfloat[4MiB Permutation with 4:1 oversubscription\label{fig:smartt_load:b}]{
    \includegraphics[width=\linewidth]{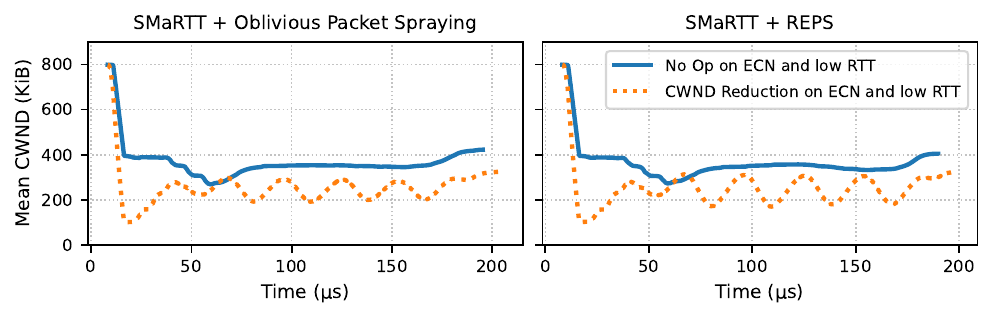}
  }
  \caption{Comparison of SMaRTT variants during a permutation workload.}
  \label{fig:smartt_load}
\end{figure}

To evaluate this decision, Figure \ref{fig:smartt_load} compares two variants of SMaRTT: in the first one we do not reduce the \textit{cwnd} in this scenario, while in the second one we do. We report the aggregated \textit{cwnd} evolution over time for all flows in a scenario without oversubscription and with oversubscription while running SMaRTT with REPS and Oblivious Packet Spraying (OPS). In the first scenario, ideally, the congestion control should ideally not activate, as this scenario can be fully solved by perfect load-balancing. Indeed, Figure \ref{fig:smartt_load:a} confirms that not reducing the window yields better performance, especially for OPS. This confirms that tightly integrating congestion control and load balancing results in better performance.

On the other hand, in Figure \ref{fig:smartt_load:b} we run the same scenario but with oversubscription, and in this case we expect the window to be reduced to match the available bandwidth. Counterintuitively, a more immediate reaction to ECN in this scenario does not necessarily yield superior performance. Our results demonstrate that our proposed design, which intentionally employs a no-operation (no-op) in this case, achieves comparable or slightly better performance. We attribute this outcome to SMaRTT's staged response and its effective integration with load balancing.

\subsubsection{Multiplicative Decrease (MD)}\label{sec:multiplicative_decrease}
The ACK is \textbf{ECN-marked} and reports an \textbf{RTT higher than the target}. This indicates a worst-case scenario and, as a consequence, SMaRTT rapidly reduces the congestion window (once per \textit{brtt}) to the difference between the target and measured average RTT, as follows:
\begin{equation}\label{eq:md}
    \begin{aligned}
        \mathit{cwnd} \mathrel{*}= \max \Big(\mathit{0.5}, \mathit{1 -} \frac{\mathit{avg\_rtt} - \mathit{trtt}}{\mathit{avg\_rtt}} \cdot \mathit{0.8}\Big)
    \end{aligned}
\end{equation}
where $\mathit{trtt}$ and $\mathit{avg\_rtt}$ are the target and measured average RTT, respectively. In any case, the multiplicative decrease reduces $\mathit{cwnd}$ at most by the half of the current $\mathit{cwnd}$. 
We use the $\mathit{avg\_rtt}$ in order to average the delay of different paths. Finally, we only execute this decrease once per $\mathit{brtt}$. \tom{$\mathit{avg\_rtt}$ is updated per-packet using an exponentially weighted moving average (EWMA).}

\subsubsection{Fair Increase (FI)}\label{sec:fair_increase}
The ACK is \textbf{not ECN-marked} but reports an \textbf{RTT higher than the target one}. This might indicate that the delay signal is outdated and the queues are not growing significantly. Consequently, SMaRTT gently increases the congestion window, ignoring the delay signal. To improve fairness (see Sec.~\ref{sec:results:incast}), the window is increased more aggressively for flows with smaller windows as follows:

\begin{equation}
    \begin{aligned}
        \mathit{cwnd} \mathrel{+}=  \frac{\mathit{p.size}}{\mathit{cwnd}} \cdot \mathit{mtu} \cdot \mathit{fi}
    \end{aligned}
\end{equation}
where $\mathit{fi}$ is a \textit{fair increase} constant and $\mathit{mtu}$ is the \textit{Maximum Transmission Unit}. The higher $\mathit{fi}$, the more aggressively the congestion window is increased. $\mathit{fi}$ impacts small flows more than large ones, and we discuss how to choose $\mathit{fi}$ in Sec.~\ref{sec:tuning}. Note, we scale the fair increase constant based on the BDP of the path. So, flows with different base RTTs (e.g., intra- vs inter- rack flows) will have different constants.
\subsubsection{Proportional Increase (PI)}\label{sec:multiplicative_increase}
The ACK is not \textbf{ECN-marked} and reports an \textbf{RTT smaller than the target one}. In this case, SMaRTT increases the congestion window proportionally to the difference between the target and measured RTTs, as follows:

\begin{equation}\label{eq:mi}
    \begin{aligned}
        \mathit{cwnd} \mathrel{+}= \min\Big(\mathit{size}, \frac{\mathit{trtt} - \mathit{p.rtt}}{\mathit{p.rtt}} \cdot \frac{\mathit{p.size}}{\mathit{cwnd}} \cdot \mathit{mtu} \cdot \mathit{pi}\Big)
    \end{aligned}
\end{equation}

$\mathit{pi}$ is a \textit{proportional increase} constant selected to increase the congestion window by at most a fixed number of MTU per RTT (depending on $\mathit{pi}$). If we assume $\mathit{cwnd}/\mathit{p.size}$ ACKs are received in an RTT, SMaRTT would increase the window size by $\frac{\mathit{trtt} - \mathit{p.rtt}}{\mathit{p.rtt}} \cdot \mathit{mtu} \cdot \mathit{pi}$ per RTT. Because we want this quantity to be smaller than an MTU, $\mathit{pi} \leq \frac{\mathit{rtt}}{\mathit{trtt} - \mathit{p.rtt}}$. Since $\mathit{p.rtt} \geq \mathit{brtt}$, where $\mathit{brtt}$ is the base RTT, we set $\mathit{pi} = \frac{\mathit{brtt}}{\mathit{trtt}-\mathit{brtt}}$. This proportional increase is followed by a fair increase to improve fairness (Sec.~\ref{sec:fair_increase}, not shown explicitly in Eq.~\ref{eq:mi}). 

\subsection{Parameter Selection}\label{sec:tuning}
\begin{description}[leftmargin=0pt]
\setlength\itemsep{1em}

\item[Window Size Bounds] SMaRTT clamps the congestion window within a predefined range. We set a maximum window to 1.5 BDP (higher than 1 to manage transient bursts) and a minimum window size of 1 MTU. \tom{Modern link speeds and delays (with a BDP around 900KiB~\cite{10154243} and 4KiB MTUs) allow incast with hundreds (200 with the mentioned parameters) of senders targeting a single receiver to be easily supported.} Higher incast degrees can be gracefully handled by SMaRTT at reduced link efficiency due to the potentially higher number of trimmed or dropped packets. Alternatively, SMaRTT can integrate a pacer-based solution similar to the one introduced by Swift for cases when the window would drop below one MTU~\cite{49448}. 

\item[ECN Marking] We test SMaRTT using RED queueing discipline, with $K_{min}$ and $K_{max}$ set to $20\%$ and $80\%$ of the switch queue size (1 $BDP$ per port) respectively~\cite{251892}.

\item[Target RTT] $\mathit{trtt}$ is selected to be equal to $1.5 \times$ the base RTT. Because we assume a queue size of 1 BDP, this corresponds to having half-full queues, thus between $K_{min}$ and $K_{max}$.



\item[Scaling] The increase constants including fair increase ($\mathit{fi}$), proportional increase ($\mathit{pi}$) change the cwnd by a constant amount every RTT independent of the cwnd (e.g., up to 1 MTU). We scale these constants based on the BDP of the network (or path). This ensures we ramp up from 0 bytes to BDP in the same wall clock time regardless of the BDP of the deployment. 
Here, the BDP can be different for different data center deployments, but also two flows within a deployment. By scaling the increase constants with the path BDPs, we ensure that even flows with different base RTTs ramp up by the same amount for the same wall clock time allowing throughput fairness for flows with different RTTs.

\end{description}

\begin{figure*}[!t]
  \centering
  \subfloat[8:1 incast\label{fig:incast:a}]{%
    \includegraphics[width=0.32\textwidth]{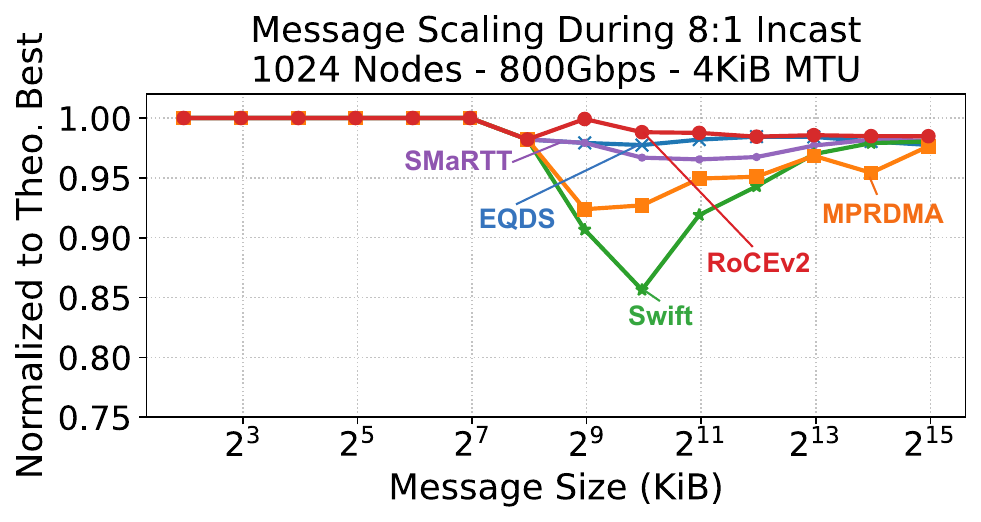}%
  }\hfill
  \subfloat[32:1 incast\label{fig:incast:b}]{%
    \includegraphics[width=0.32\textwidth]{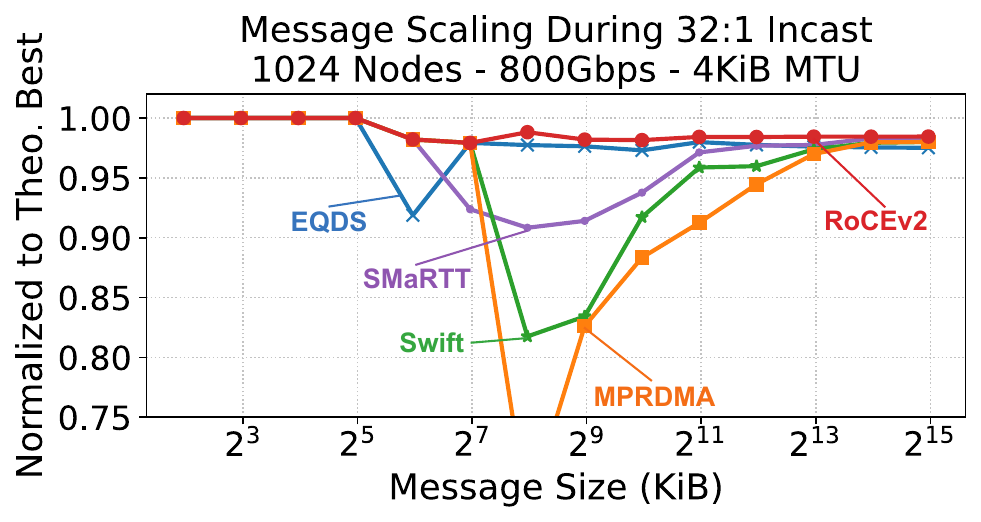}%
  }\hfill
  \subfloat[50:1 incast\label{fig:incast:c}]{%
    \includegraphics[width=0.32\textwidth]{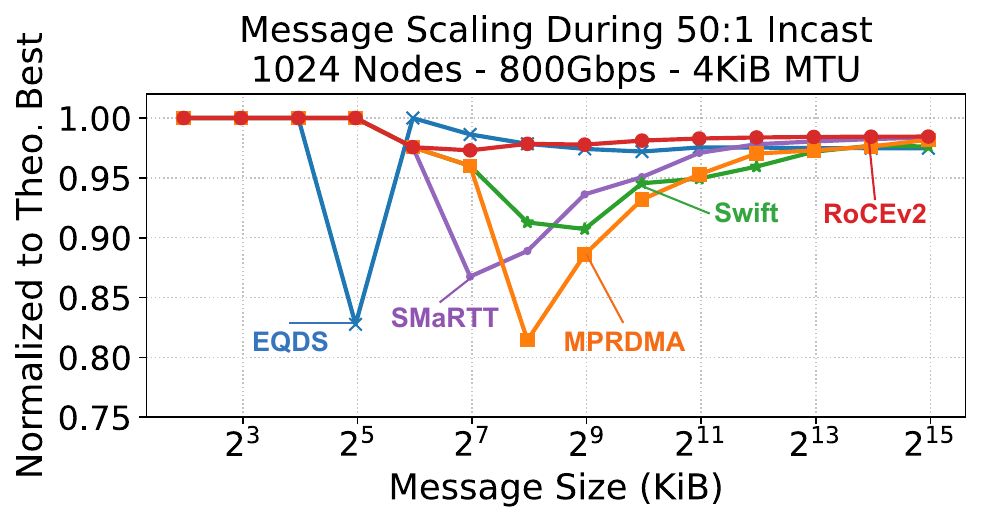}%
  }
  \caption{Relative incast performance for different message sizes and incast degrees. The $y$-axis number indicates the relative performance compared to what is theoretically achievable with perfect congestion control.}
  \label{fig:incast_f}
\end{figure*}

\section{Simulation Evaluation}
\label{sec:results}
\paragraph{Simulation Setup} We ran simulations using and extending the \textit{htsim} packet-level network simulator~\cite{10.1145/3098822.3098825}. Our simulations consider four different fat-tree topologies, two non-blocking (one with 1,024 and the other with 128 nodes), and three with 1,024 nodes and 2:1, 4:1, and 8:1 oversubscription ratios, due to the relevance of such topologies in production datacenters~\cite{43837}. We set a 4KiB MTU size, a 800 Gbps network bandwidth, and a 400ns switch traversal latency, consistent with the latest generation of switches~\cite{toma5,9355230}. For the sake of simplicity, we assume that all links have the same length and, hence, the same latency, equal to 600ns. 

\paragraph{State-of-the-art Comparison} \tom{We evaluate SMaRTT against EQDS (which uses a NDP-derived control loop), Swift, MPRDMA, RoCEv2. To have a fair comparison and focus on congestion control, we enable trimming for all the algorithms as loss detection signal.} 
\paragraph{Analyzed Workloads} We consider the following three traffic patterns: (i) \textbf{incast}: relevant for traditional datacenter workloads, that often send requests to a large number of workers, and must then handle their near-simultaneous responses; (ii) \textbf{permutation}: simulates point to point connections between pairs of nodes, selected so that each packet crosses the core switches, to emulate a worst case scenario and stress the load balancer; (iii) \textbf{alltoall}: relevant for many AI workloads such as Mixture of Experts models. We note that we use these workloads also for the hardware evaluation in Section~\ref{sec:hw_results}.

\subsection{Incast}\label{sec:results:incast}
Fig.~\ref{fig:incast_f} shows the relative performance of SMaRTT in incast scenarios ranging from an 8:1 incast to a 50:1 incast for varying message sizes. We note that EQDS, being a receiver-based mechanism, can easily achieve good performance. Indeed, it can precisely communicate the window size to each sender and schedule them to be perfectly fair. 

On the other hand, sender-based mechanisms like MPRDMA, SMaRTT, and Swift perform slightly worse (beside the lossless RoCEv2), particularly true for medium sized messages. Small messages do not overrun switch buffers, whereas for large enough messages all the algorithms have enough time to converge to the correct rate. The zone in between, usually for messages slightly smaller than BDP, is subject to random drop events that affect sender-based mechanisms more. Moreover, MPRDMA seems to have slightly worse performance due to its lack of fairness features.

\subsection{Permutation}\label{sec:results:permutation}
\begin{figure*}[!t]
  \centering
  \subfloat[Permutation 1:1 OS - 32MB\label{fig:permsum:a}]{%
    \includegraphics[width=0.24\textwidth]{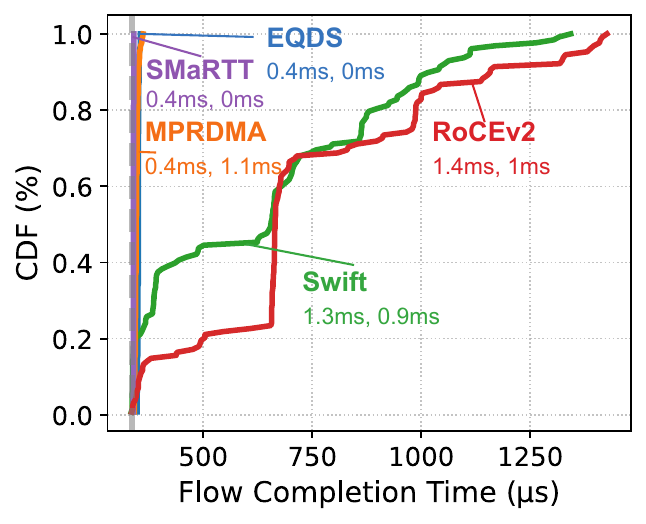}%
  }\hfill
  \subfloat[Permutation 2:1 OS - 32MB\label{fig:permsum:b}]{%
    \includegraphics[width=0.24\textwidth]{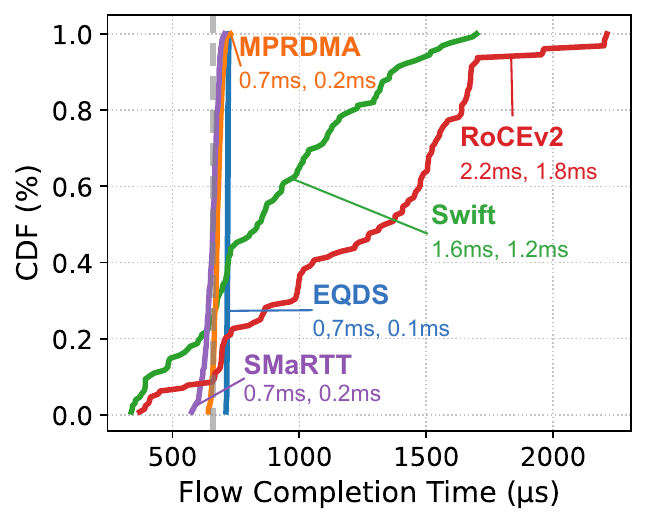}%
  }\hfill
  \subfloat[Permutation 4:1 OS - 32MB\label{fig:permsum:c}]{%
    \includegraphics[width=0.24\textwidth]{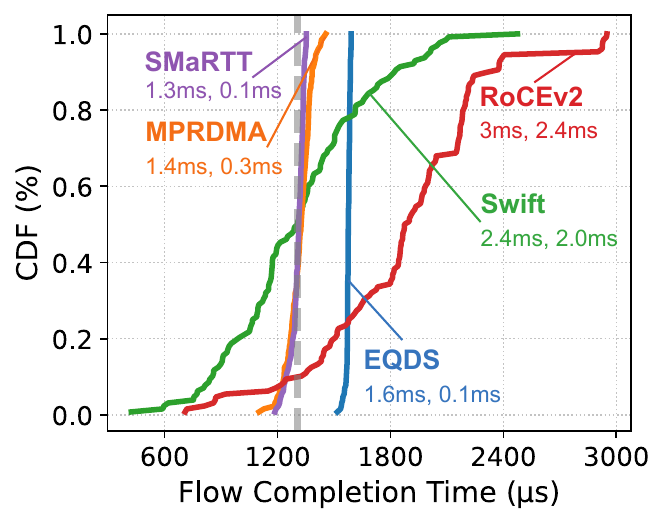}%
  }\hfill
  \subfloat[Uneven Perm. 8:1 OS - 32MB\label{fig:permsum:d}]{%
    \includegraphics[width=0.24\textwidth]{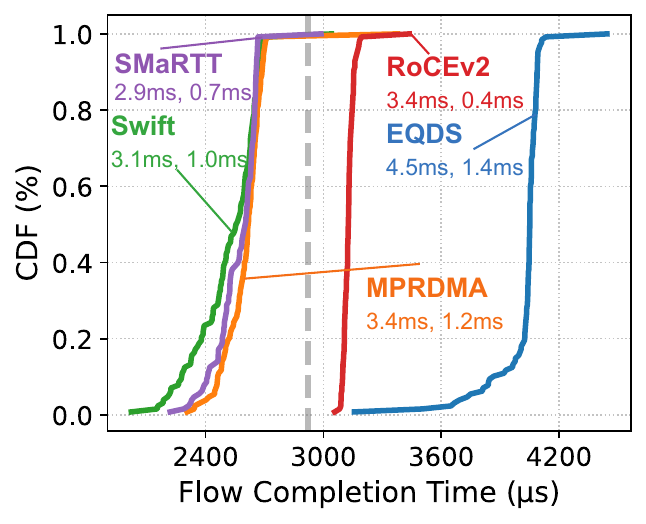}%
  }
  \caption{Flow completion time with four different permutations on fat trees with different oversubscription ratios.}
  \label{fig:permsum}
  \vspace{-1mm}
\end{figure*}

In Fig.~\ref{fig:posterchild}, we show the results for a 2 MiB and 32 MiB permutation traffic pattern on an 8:1 oversubscribed fat tree with 1,024 nodes. It is worth noting that, when using a relatively small message of 2 MiB, just slightly bigger than the BDP, SMaRTT outperforms all other algorithms while being fair (it can quickly react thanks to ECN). In Fig.~\ref{fig:permsum:a}, Fig.~\ref{fig:permsum:b}, and Fig.~\ref{fig:permsum:c} we show what happens on 1:1, 2:1, and 4:1 oversubscribed fat trees respectively. In these cases, SMaRTT still outperforms the other algorithms, but not as much as before as a smaller oversubscription ratio is easier to manage. 

Finally, we show in Fig.~\ref{fig:permsum:d} a scenario with 32MiB flows, except for one that is bigger and sends 64MiB. This scenario shows how SMaRTT outperforms other sender-based mechanisms by almost 30\% thanks to its capacity to reclaim bandwidth quickly (Sec.~\ref{sec:fast_increase}).

Generally, we can notice that for high oversubscription ratios, the performance of EQDS degrades even more due to the high number of trimmed packets. To emphasize this point, we measure the number of trimmed packets for EQDS and report that, in the worst case scenario just mentioned, it can generate up to 155x more trimmed packets than SMaRTT, resulting in wasted bandwidth and resources. The EQDS authors state that in such cases a CC algorithm on the sender side would benefit its performance, and we explore that in Sec.~\ref{sec:results:augmenting}.

\subsection{Alltoall}\label{sec:results:alltoall}
\begin{figure}[hbtp]
        \begin{center}
            \includegraphics[width=0.85\columnwidth]{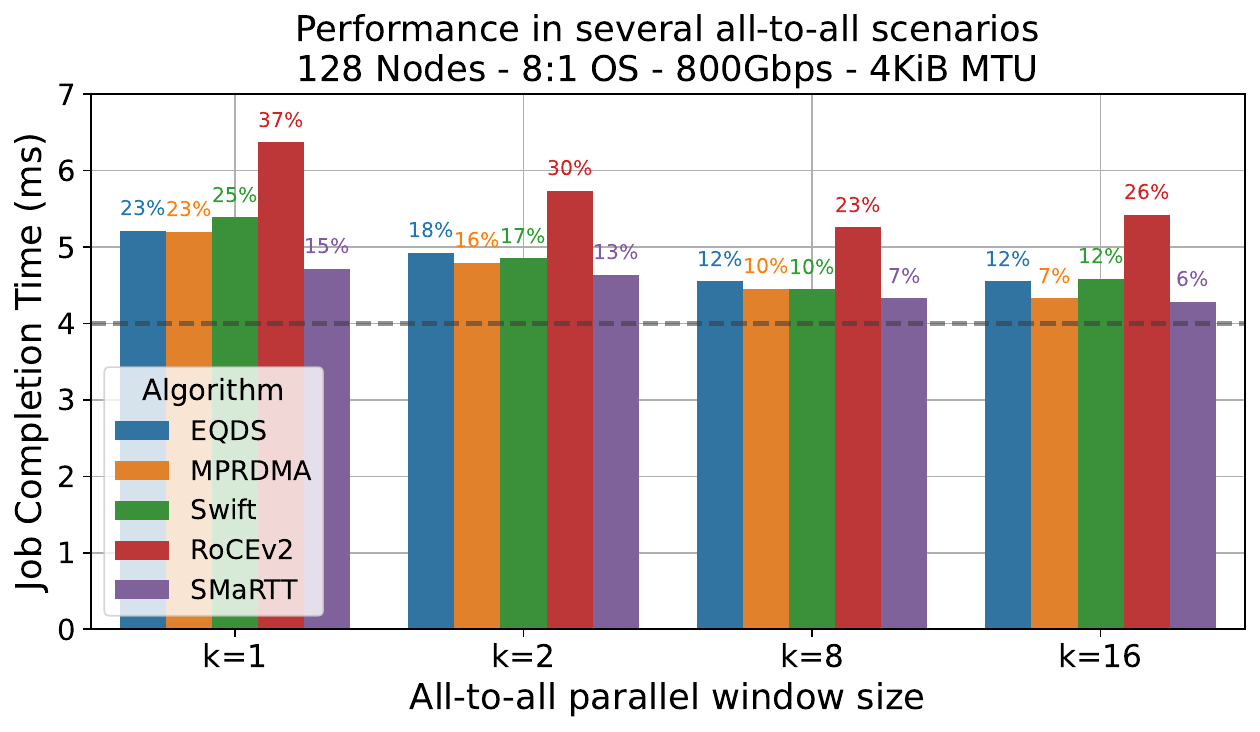}
        \end{center}
        \caption{The number at the top of each bar indicates the time delta from ideal completion time (grey dotted line). \textit{k} indicates the max number of active flows for a sender.}
        \label{img:alltoall}
        \vspace{-1mm}
\end{figure}

We also evaluate performance for alltoall communication, a collective operation commonly used in AI workloads~\cite{10.5555/3571885.3571899}. Due to the large number of messages generated, we use the smaller topology to reduce simulation times. We use a windowed algorithm for the alltoall to have at least $k$ active flows per node at any time. We show the results in an oversubscribed fat tree, as a non-blocking tree topology can theoretically handle alltoall traffic without problems for congestion control.
We report the results of our evaluation in Fig.~\ref{img:alltoall}. In all scenarios SMaRTT takes the lead being only up to 6\% behind the ideal time and up to 20\% better than other sender-based schemes.

 \vspace{-1mm}

\section{Hardware Evaluation}
\label{sec:hw_results}
The simulation evaluation shows SMaRTT is nearly optimal across a range of traffic patterns, but the simulator models a perfect stack implementation that does not add any delay to the packets during processing. \textit{Does SMaRTT behave similarly in a real network, running on real hardware?}
This question is particularly important as SMaRTT relies on a combination of multiple algorithms (QuickAdapt, Fast Increase, etc), which are naturally more expensive computationally than traditional congestion control (e.g., DCTCP),  and on measurements of the queueing delay which will be noisier in practice.

We have implemented SMaRTT using the DPDK framework. The implementation uses one CPU core which runs in a continuous loop and in each iteration: (1) takes packets from receiver ring of the application / upper transport stack, processes them and sends them out on the network interface if congestion control state allows it, or buffers them locally otherwise; (2) processes incoming packets from the network, which can be: (a) ACKs / NACKs which are used to update packet state (e.g. queue for retransmit) and to run the congestion control loop, including sending new packets when allowed. (b) incoming data or trimmed packets, for which it generates ACKs or NACKs as appropriate; data packets are decapsulated and sent to the local upper layer stack via the local transmit ring; (3) checks the transmit queue to decide which packets have timed out and thus need retransmitting. 

We have implemented a libfabric \cite{libfabric} provider which talks to our SMaRTT stack via software rings. We then test the congestion control using a libfabric test application which takes a connection matrix file as input that specifies all connections and eventual dependencies. 

\paragraph{Testing Setup} We use a small-scale testbed to test our application. We use two Broadcom Trident 4 switches with trimming support to implement a leaf-spine topology with 4 top-of-rack and 4 spine switches; each TOR switch is connected to each of the spine switches, and has four interfaces to hosts.  Each emulated switch uses a separate virtual-routing function (VRF) which has its own routing table. Connections via VRFs are done using DAC cables. Our hosts are Supermicro servers with AMD EPYC processors, 128GB of RAM, and use Broadcom Thor 2x100Gbps NICs. 

Our implementation can sustain 100Gbps unidirectional throughput, but software overheads prevent it from reaching bidirectional 100Gbps performance (150Gbps in total). To measure the quality of the congestion controller rather than the software overheads, we reduce the linkspeeds to 50Gbps in our testbed and run all tests comparing SMaRTT and RoCEv2 using PFC.

\subsection{Incast Results (DPDK)}

\begin{figure}
    \includegraphics[width=0.85\columnwidth]{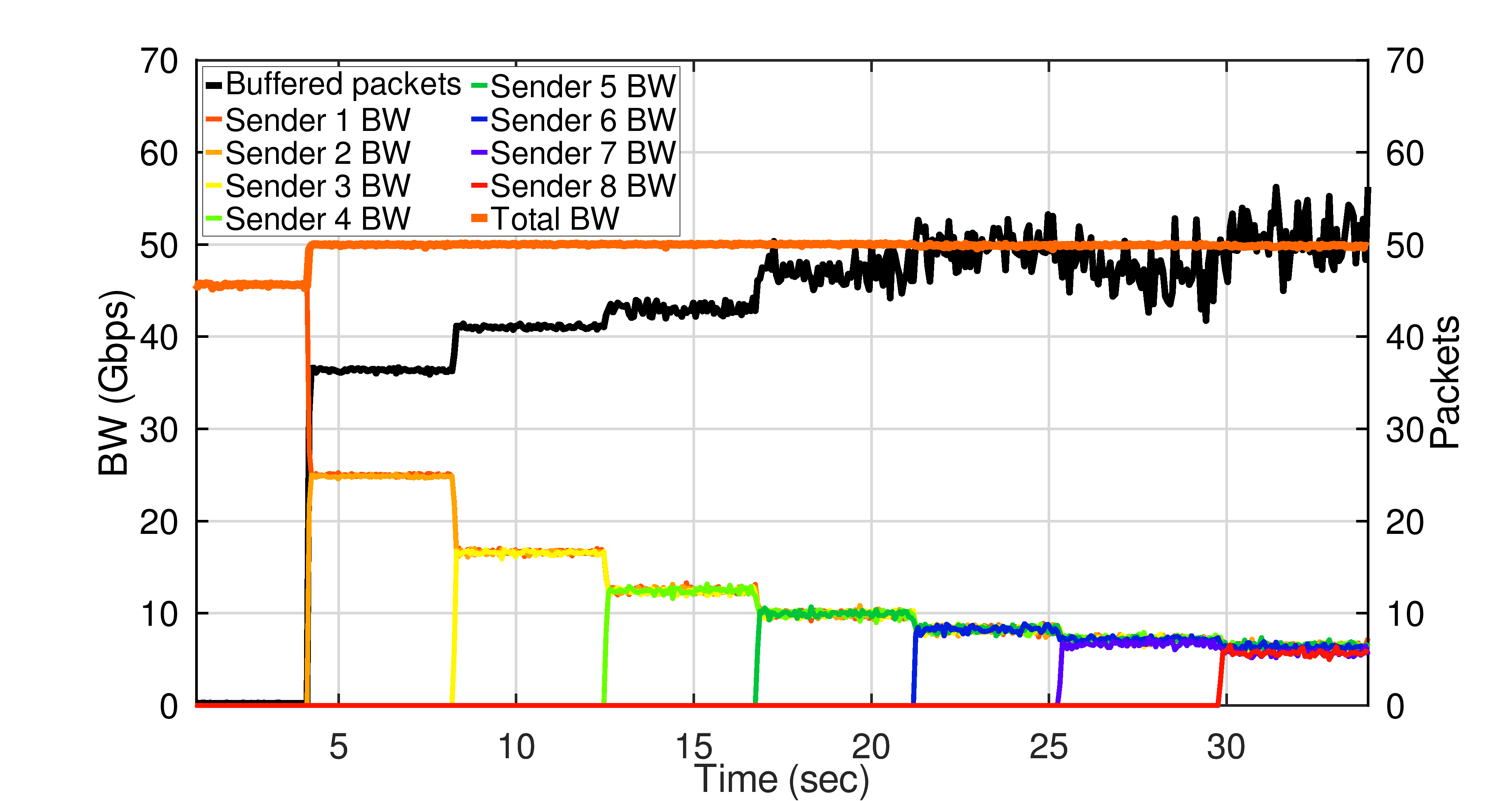}
    \caption{Incast deep-dive (DPDK implementation)}
    \label{fig:incast_longflow}
    \vspace{-0.7em}
\end{figure}

\begin{figure*}[!t]
  \centering
  \subfloat[Incast FCT inflation vs.\ ideal\label{fig:incast_dpdk}]{%
    \includegraphics[width=0.32\textwidth]{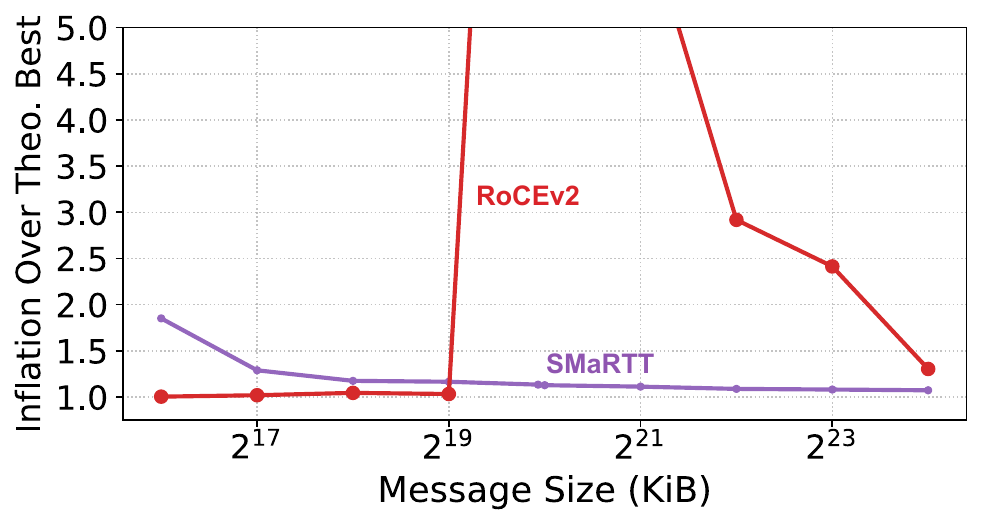}%
  }\hfill
  \subfloat[Permutation FCT inflation\label{fig:permutation_dpdk}]{%
    \includegraphics[width=0.32\textwidth]{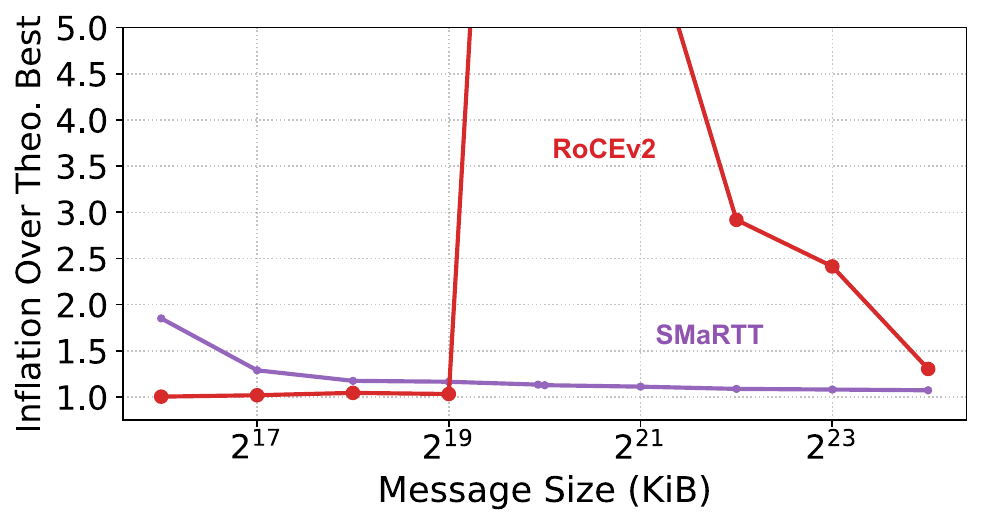}%
  }\hfill
  \subfloat[Sequential all-to-all FCT inflation\label{fig:alltoall_seq}]{%
    \includegraphics[width=0.32\textwidth]{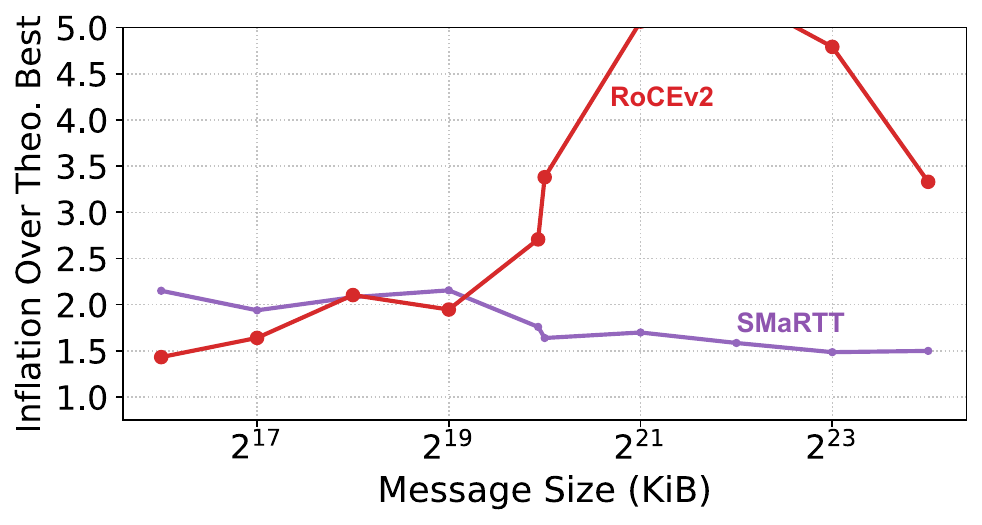}%
  }
  \caption{Experiments comparing SMaRTT DPDK to hardware RoCEv2 on the hardware testbed.}
\end{figure*}

In our first experiment, we run an incast with long-running flows that arrive every \~5s. Figure \ref{fig:incast_longflow} plots per-flow throughput, as well as the throughput of the bottleneck link and the bottleneck queue size (shown with a black line, Y2 axis). The result confirms that SMaRTT shares capacity fairly across the active flows and it is able to control the queue size to match the target delay, despite inherently noisy RTT measurements in our software stack. 

Next, we use the libfabric test app and run a 15-to-1 incast where all flows are the same size and start at the same time. Figure \ref{fig:incast_dpdk} shows the flow completion time inflation compared to the theoretical optimal, for both SMaRTT and RoCEv2. Note that RoCE has optimal performance for very small flows (smaller than 256KB) because there are no network effects to mention; however, for larger flows DCQCN congestion control kicks in and results in underutilization of the link. Finally, SMaRTT has near-optimal performance for large flows, and is slightly slower for small flows because the added latency of the software implementation (a few us) is comparable to the optimal flow completion times.

Note that, with finer tuning, RoCEv2 can get near-optimal performance for this traffic pattern,
but tailoring the parameters for one traffic pattern affects performance in others. Hence, we chose to run with the default parameters that give steady performance for the all-to-all implementation.

\subsection{Permutation Results (DPDK)}

Figure \ref{fig:permutation_dpdk} compares the flow completion time of a 16-node permutation to the
theoretical optimal in our testbed, as we vary the flow size. The results for SMaRTT are very similar to
the incast, where it is roughly twice slower than optimal for small flows, and within 20\% of optimal for
flows larger than 4MB; these long flow results match our simulation experiments, giving us confidence 
that SMaRTT will achieve predicted performance in deployment. In contrast, RoCE flow completion times are comparable for small flows, but they fall off a cliff at 1MB-2MB due to DCQCN overreaction and ECMP collisions, and are 3-5 times slower than optimal for larger flows, mostly due to flow collisions. 
 
\subsection{All-to-all Results (DPDK)}

Figure \ref{fig:alltoall_seq} shows the flow completion time of a sequential all-to-all implementation as we vary the flowsize. In this implementation, nodes are organized in a logical ring, and there are n-1 iterations; in the first iteration, a node sends to its immediate neighbor; in the second iteration to its second neighbor to the right, and so on. This is, in effect, a sequence of permutations. For small flows our software implementation overheads make SMaRTT performance be similar to RoCE (around 2x slower than optimal); for larger flows SMaRTT is within 1.5x of optimal, while RoCEv2 is 3-6x slower. This matches our simulation results, provides confidence that the simulation results will be reproduced in deployment.
 
\section{Augmenting EQDS with SMaRTT}\label{sec:results:augmenting}
One of the main drawbacks of using a receiver-based mechanism like EQDS, as shown in the previous results, is its bad management of fabric congestion. As suggested in the EQDS paper, it should be possible to improve EQDS' performance by complementing it with a sender-based CC algorithm. To do so, we explore using EQDS with SMaRTT, as well as with a DCTCP-style CC algorithm (MPRDMA). While running in combination with EQDS, the role of the sender-based congestion control is to limit its sending rate by capping the congestion window and adapting as needed.     

In theory, this combination should give us the benefits of both worlds: perfect incast management thanks to the receiver-based CC and the ability to deal with the other forms of congestion thanks to a solid sender-based CC. In this work, we only show a simple scenario where such combinations seem beneficial for EQDS. A more in-depth analysis of such a combination is necessary but outside the scope of this paper.
\begin{figure}[!t]
  \centering
  \subfloat[All flows 2MiB.\label{fig:augmenteqds:a}]{%
    \includegraphics[width=0.49\linewidth]{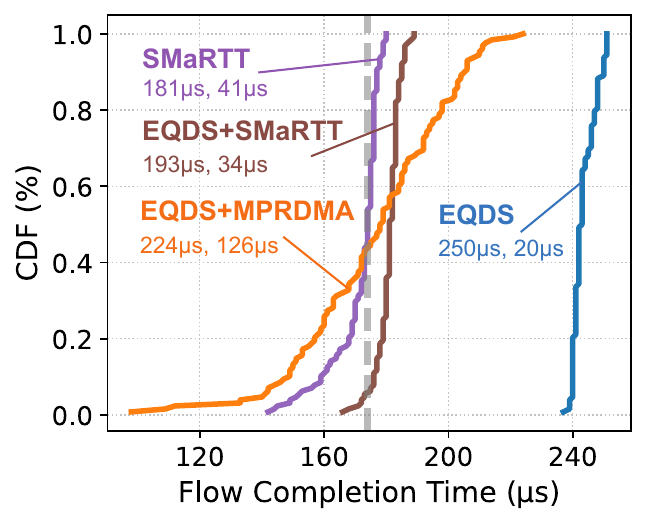}%
  }\hfill
  \subfloat[One longer flow at 64 MiB.\label{fig:augmenteqds:b}]{%
    \includegraphics[width=0.49\linewidth]{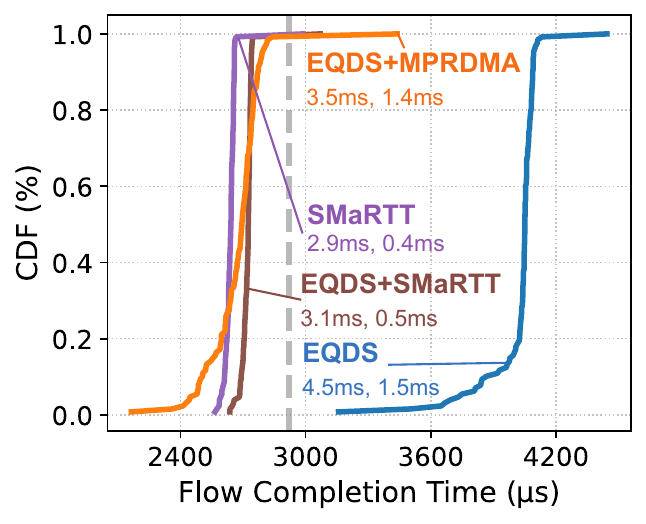}%
  }
  \caption{Augmenting EQDS with sender-based mechanisms.}
  \label{img:augmenteqds}
  \vspace{-1mm}
\end{figure}
In Fig.~\ref{img:augmenteqds} we report a permutation scenario on an 8:1 oversubscribed fat tree, with 2 MiB and 32 MiB flows. We can see that SMaRTT improves EQDS performance. For 2MiB flows, the improvement introduced by SMaRTT is higher than that of MPRDMA, thanks to its better fairness. When running with 32 MiB flows except for one being 64 MiB, we can also see how we do very well on our own but also help EQDS. In this case having FastIncrease is key to recover quickly the bandwidth for the larger flow.
\section{Related Work}
Researchers proposed several datacenter CC algorithms, and we discuss in the following how SMaRTT differs from some of the most recently proposed ones. Receiver-based algorithms such as NDP~\cite{10.1145/3098822.3098825}, EQDS~\cite{278348}, pHost~\cite{10.1145/2716281.2836086}, ExpressPass~\cite{9047463}, SMSRP~\cite{10.1145/2807591.2807600}, and Homa~\cite{10.1145/3230543.3230564} maintain end-to-end credits that are used to adjust sender flow rates. They can effectively manage incast scenarios by precisely regulating the transmission rates of the senders. We have shown, however, that they can struggle when dealing with fabric congestion.

On the other hand, sender-based CC algorithms can deal better with fabric congestion, but might not be as responsive when confronted with sudden workload shift (e.g. incast or dynamic job arrivals or departures). Some of those algorithms like DCTCP~\cite{10.1145/1851182.1851192}, Hull~\cite{180607}, and D$^2$TCP~\cite{10.1145/2342356.2342388} rely on ECN marking. For example, DCQCN~\cite{zhu2015congestion} combines ECN and \textit{Priority-based Flow Control} (PFC) to avoid packet losses and react to congestion quickly. However, PFC is hard to tune and can cause PFC storms~\cite{10154243,10.1145/3230543.3230557}. Differently from DCQCN, SMaRTT is designed for lossy networks and relies on packet trimming to avoid switches dropping packets, thus avoiding all the issues related to the use of PFC. Other sender-based algorithms Swift~\cite{49448} and TIMELY~\cite{43840} detect congestion using delay but, as we show in this paper, they cannot react as fast as ECN-based algorithms.


Lastly, some work recently demonstrated that having an unfair CC can improve the performance of jobs that run AI training workloads~\cite{10.1145/3563766.3564115}. These works are orthogonal to SMaRTT and target fairness between different jobs.

\vspace{-2.5mm}
\section{Conclusion}
This paper proposes SMaRTT, a sender- and window-based congestion control algorithm using ECN and delay as congestion signals and packet trimming to detect packet losses. When trimming is not supported, SMaRTT falls back to a timeout-based mechanism, which we show adds at most one RTO worth of delay. We evaluate SMaRTT on several workloads and topologies, showing that it consistently matches or outperforms the performance of EQDS, Swift, RoCEv2, and MPRDMA by up to 50\%. We also demonstrate how SMaRTT can complement receiver-based CC algorithms, such as EQDS, by helping address in-network congestion.

\section{Acknowledgments}
The authors thank Michael Papamichael, Adrian Caulfield, and Yanfang Le for their valuable feedback and help. This work is supported by the European Union’s Horizon Europe under grant 101175702 (NET4EXA), the Sapienza University Grants ADAGIO and D2QNeT (Bando per la ricerca di Ateneo 2023 and 2024), the European Research Council (ERC) under the European Union’s Horizon 2020 research and innovation program (grant agreement PSAP, No. 101002047). We also thank the Swiss National Supercomputing Center (CSCS) for providing the computational resources used in this work. ChatGPT-5 assisted with editing and quality control.

\bibliographystyle{IEEEtran}
\bibliography{bib}

\end{document}